\documentclass[12pt]{article}

\usepackage{amsmath,amssymb,graphicx} 
\usepackage{epsf}
\newcommand{\feyn}[4]{\hline $#1$ & $#2$ & $#3$ & $#4$ \\}
\newcommand{\beq}{\begin{eqnarray}}
\newcommand{\eeq}{\end{eqnarray}}
\newcommand{\arline}{\nonumber \\}
\newcommand{\met}{g_{\mu \nu}}
\newcommand{\centeron}[2]{{\setbox0=\hbox{#1}\setbox1=\hbox{#2}\ifdim
                           \wd1>\wd0\kern.5\wd1\kern-.5\wd0\fi \copy0
                           \kern-.5\wd0\kern-.5\wd1\copy1\ifdim\wd0>\wd1
                           \kern.5\wd0\kern-.5\wd1\fi}}
\newcommand{\ltap}{\>\centeron{\raise.35ex\hbox{$<$}}
                   {\lower.65ex\hbox{$\sim$}}\>}
\newcommand{\gtap}{\>\centeron{\raise.35ex\hbox{$>$}}
                   {\lower.65ex\hbox{$\sim$}}\>}
\newcommand{\gsim}{\mathrel{\gtap}}
\newcommand{\lsim}{\mathrel{\ltap}}

\newcommand\ZZ{\hbox{\zfont Z\kern-.4emZ}}
\font\zfont = cmss10 

\newcommand{\EPS}{\frac{v}{f}}

\newcommand{\whp}[1]{W_H^{+#1}}
\newcommand{\whm}[1]{W_H^{-#1}}
\newcommand{\ah}[1]{A_H^{#1}}
\newcommand{\wlp}[1]{W^{+#1}}
\newcommand{\wlm}[1]{W^{-#1}}
\newcommand{\zh}[1]{Z_H^{#1}}
\newcommand{\cw}{c_w}
\newcommand{\sw}{s_w}
\newcommand{\phs}{\phi^P}
\newcommand{\php}{\phi^+}
\newcommand{\phm}{\phi^-}
\newcommand{\pho}{\phi^0}
\newcommand{\phpp}{\phi^{++}}
\newcommand{\phmm}{\phi^{--}}
\newcommand{\moms}{\left( p_1 - 2 p_2 \right)_\mu}
\newcommand{\tpb}{\bar{t}'_+}
\newcommand{\tm}{t'_-}
\newcommand{\tmb}{\bar{t}'_-}
\newcommand{\tp}{t'_+}
\newcommand{\EPSs}{\frac{v^2}{f^2}}

\def\be{\begin{equation}}
\def\ee{\end{equation}}
\def\bea{\begin{eqnarray}}
\def\eea{\end{eqnarray}}
\def\lsim{\mathrel{\raise.3ex\hbox{$<$\kern-.75em\lower1ex\hbox{$\sim$}}}}
\def\gsim{\mathrel{\raise.3ex\hbox{$>$\kern-.75em\lower1ex\hbox{$\sim$}}}}
\def\ifmath#1{\relax\ifmmode #1\else $#1$\fi}
\begin{document}
\begin{titlepage}
\begin{flushright}
{\tt hep-ph/0411264} \\
\end{flushright}

\vskip.5cm
\begin{center}
{\huge \bf Phenomenology of the Littlest Higgs with T-Parity}
\vskip.2cm
\end{center}

\begin{center}
{\bf {Jay Hubisz and Patrick Meade}} \\
\end{center}
\vskip 8pt

\begin{center}
{\it Institute for High Energy Phenomenology, \\
F.R. Newman Laboratory of Elementary Particle Physics, \\
Cornell University, Ithaca, NY 14853, USA } \\
\vspace*{0.3cm}{\tt  hubisz@mail.lepp.cornell.edu,
meade@mail.lepp.cornell.edu}
\end{center}

\vglue 0.3truecm

\begin{abstract}
\vskip 3pt \noindent

Little Higgs models offer an interesting approach to weakly
coupled electroweak symmetry breaking without fine tuning. The
original little Higgs models were plagued by strong constraints
from electroweak precision data which required a fine tuning to be
reintroduced.  An economical solution to this problem is to
introduce a discrete symmetry (analogous to R-parity of SUSY)
called T-parity. T-parity not only eliminates most constraints
from electroweak precision data, but it also leads to a promising
dark matter candidate.  In this paper we investigate the dark
matter candidate in the littlest Higgs model with T-parity.  We
find bounds on the symmetry breaking scale $f$ as a function of
the Higgs mass by calculating the relic density. We begin the
study of the LHC phenomenology of the littlest Higgs model with
T-parity. We find that the model offers an interesting collider
signature that has a generic missing energy signal which could
``fake" SUSY at the LHC. We also investigate the properties of the
heavy partner of the top quark which is common to all littlest
Higgs models, and how its properties are modified with the
introduction of T-parity.  We include an appendix with a list of
Feynman rules specific to the littlest Higgs with T-parity to
facilitate further study.

\end{abstract}

\end{titlepage}

\newpage


\section{Introduction}
\label{intro} \setcounter{equation}{0} \setcounter{footnote}{0}

Within the next decade the mechanism responsible for electroweak
symmetry breaking (EWSB) will hopefully be revealed by the LHC.
For theoretical physicists this is a time at which many attempt to
conjecture every possibility before experimental data ultimately
chooses the correct one (or at least narrows the list). The
standard paradigm for weakly coupled EWSB is the Higgs mechanism,
however in the Standard Model (SM) there are quadratically
divergent diagrams which contribute to the Higgs mass.  If new
physics to cut off the quadratic divergences to the Higgs mass
does not occur at approximately the TeV scale, the SM will be a
finely tuned theory.  As long as one takes naturalness as a guide
for model building this implies that there should be new physics
at the TeV scale associated with the physics of EWSB. For two
decades the standard for new physics at the TEV scale, which
cancels the quadratic divergences of the SM Higgs, was
supersymmetry.  With the turn on of the LHC drawing near there
have recently been many attempts at coming up with viable
alternatives to supersymmetry.  The alternative that will be
focused on in this paper will be the little Higgs
mechanism~\cite{little1}.

The origin of the little Higgs idea dates back almost as far as
supersymmetry to the papers of Georgi et
al.~\cite{HiggsPseudo,KaplanGeorgi} that attempted to realize the
Higgs as a pseudo-Goldstone boson (PGB). These original papers
were unsuccessful due to the fact that they reintroduced a fine
tuning to keep the symmetry breaking scale that generates the
Goldstone separate from the electroweak scale. The new mechanism
that makes the Higgs ``little" in the current reincarnation of the
PGB idea is collective symmetry breaking~\cite{little1}.
Collective symmetry breaking protects the Higgs by several
symmetries under each of which the Higgs is an exact Goldstone.
Only if the symmetries are broken collectively, i.e. by more than
one coupling in the theory, can the Higgs pick up a contribution
to its mass and hence all one loop quadratic divergences (which
involve one coupling alone) to the Higgs mass are avoided.

The generic structure of little Higgs
models~\cite{littlest,littlestmoose,witek,martin1,cust1,terning,cust2,martin2}
is a global symmetry broken at a scale $f$ which is around a TeV.
At the scale $f$ there are new gauge bosons, scalars, and fermions
responsible for cancelling the one loop quadratic divergences to
the Higgs mass from SM particles (for a brief review of the gauge
and global symmetries of most little Higgs models
see~\cite{susyproc}).  Even though there are no direct
experimental signatures that guide model building beyond the SM,
there are many indirect constraints that a model of new physics
must satisfy. Whenever one tries to introduce new particles around
the TeV scale which couple to SM particles there is usually a
tension with precision EW measurements which typically favor the
scale of new physics to be $\sim\,5\,$-$10$ TeV.  This tension
between having to introduce new physics at the TeV scale for
naturalness, and EW precision tests (EWPT) preferring the scale of
new physics to be a factor of $\sim 10$ higher is the so called
``little" hierarchy problem. The original little Higgs models
unfortunately did not relieve the tension caused by EWPT. The
scale $f$ had to be raised significantly above a TeV, which
reintroduced a fine tuning to the Higgs mass~\cite{us1,slac,us2}.
New little Higgs models~\cite{cust1,cust2} were introduced with
much larger symmetry structures that could incorporate a custodial
$SU(2)$ symmetry which then ameliorated most of the problems from
EW constraints.  In keeping with the idea of looking for
alternatives to SUSY a more economical solution to the ``little"
hierarchy problem was proposed in~\cite{wudka,tpar1}.

In trying to find alternatives to the MSSM one should examine the
model critically to find what are the keys to its success.  The
MSSM is unarguably the most theoretically well motivated extension
of the SM, including a stable hierarchy between the weak and
Planck scales, as well as gauge coupling unification.  However the
key to the MSSM is $R$-parity, a discrete symmetry introduced by
hand, without which the MSSM would be ruled out by experiment. The
motivation for $R$-parity is normally to forbid operators that
would lead to rapid proton decay, but with the introduction of
$R$-parity it also forbids dangerous four-fermion operators and
contributions to Z-pole observables.  The $R$-parity also governs
the signatures of the MSSM in colliders since the conservation of
$R$-parity requires all $R$-odd particles to be pair produced. If
the lightest parity odd particle in the MSSM is neutral, the
signal in detectors will be missing energy.

If the leading candidate for new physics at the TeV scale relies
upon a discrete symmetry to be phenomenologically acceptable it is
quite plausible that its alternatives could incorporate a discrete
symmetry as well~\cite{wudka,tpar1}.  Introducing a discrete
symmetry called ``T-parity" into little Higgs models was done
first in~\cite{tpar1}, and applied to other little Higgs models
in~\cite{tpar2,tpar3}.  T-parity is a natural symmetry of most
little Higgs models where SM particles are even under the symmetry
while most of the new particles at the TeV scale are odd.  Most of
the constraints from EWPT on little Higgs
models~\cite{us1,slac,us2} come from tree level mixing of heavy
and light mass eigenstates which T-parity forbids.  T-parity
therefore solves the little hierarchy problem for the models in
which it can be implemented~\cite{tpar2}. T-parity also has a
further benefit, if the lightest parity odd particle is neutral
and T-parity is conserved it will be a candidate for a dark matter
WIMP (just as R-parity provides a candidate WIMP under the same
circumstances for supersymmetry).

In this paper we will study the phenomenology of the littlest
Higgs model~\cite{littlest} with the addition of
T-parity~\cite{tpar2,tpar3}.  There have been many studies of the
phenomenology of the littlest Higgs
model~\cite{maxim1,han1,maxim2}; however with the introduction of
T-parity most of these studies do not apply since the T-odd
particles can not be singly produced.  We begin with reviewing the
littlest Higgs model with T-parity in
Section~\ref{sec:model}.   It turns out that there are several
different approaches to incorporating T-parity into the littlest Higgs
model and we discuss in Section~\ref{alternatives} how most
reduce to a similar low energy effective spectrum that we study. In
Section~\ref{darkmatter} we investigate the phenomenology of the
dark matter candidate in the littlest Higgs model which is a heavy
neutral gauge boson.  In Section~\ref{collider} we explore the
discovery possibilities for the littlest Higgs with T-parity at
the LHC.  In Appendix~\ref{app:fourfermi}, we discuss four fermion
operators in the model that we chose to study.  In
Appendix~\ref{app:feyn} we list Feynman rules for the littlest
Higgs with T-parity to facilitate further study of the model.

\section{Review of the ``Littlest'' Higgs Model with T-Parity} \label{sec:model} \setcounter{equation}{0}

There are currently three known ways to implement T-parity in the
littlest Higgs model~\cite{tpar2,tpar3}.  The first approach uses
the technology of Callan, Coleman, Wess, and Zumino
(CCWZ)~\cite{tpar2,cwz}, where the model is almost identical to
the original littlest Higgs model except for the fermion sector.
The other approaches~\cite{tpar3} to implementing T-parity in the
littlest Higgs rely on enlarging the global and gauge symmetry
structure but are similar to the CCWZ littlest Higgs model with T-parity in the low energy
limit. For the sake of studying a concrete model we will first
discuss in detail the CCWZ approach that the calculations in this
paper are based upon. We will then discuss the other ways to
implement T-parity and how the low energy phenomenology of all
three implementations are approximately the same.

\subsection{Littlest Higgs with T-parity}\label{subsec:tparity}
The original littlest Higgs model was based on a non-linear
$\sigma$ model describing an \break $SU(5)/SO(5)$ symmetry
breaking~\cite{littlest} and serves as the starting point for
including T-parity~\cite{tpar2}. The symmetry breaking can be
thought of as originating from a vacuum expectation value(VEV) of
a symmetric tensor of the $SU(5)$ global symmetry. A convenient
basis for this breaking is characterized by the direction
$\Sigma_0$ for the VEV of the form
\begin{equation}
\Sigma_0 =\left( \begin{array}{ccccc} &&&1& \\&&&&1\\ &&1\\1\\&1 \end{array}
\right).
\end{equation}
The Lie algebra made up of the broken generators, $X^a$, and
unbroken generators, $T^a$, has an automorphism where $T^a\to T^a$
and $X^a\to -X^a$, which can be expressed as $\tau^a\to -\Sigma_0
(\tau^a)^T \Sigma_0$ for any generator $\tau_a$.  This
$Z_2$ automorphism of the symmetric space $SU(5)/SO(5)$ will be a useful guide in
implementing T-parity.  The Goldstone fluctuations are described
by the pion fields $\Pi = \pi^a X^a$. The non-linear sigma model
field is then
\begin{equation}
\Sigma (x) = e^{i\Pi/f} \Sigma_0 e^{i \Pi^T/f}=e^{2i\Pi/f} \Sigma_0.
\end{equation}
where $f$ is the value of the VEV that accomplishes the breaking.
An $[SU(2)\times U(1)]^2$ subgroup of the SU(5) global symmetry is
gauged, where the generators of the gauged symmetries are given by
\begin{eqnarray}\label{gauged}
&Q_1^a=\left( \begin{array}{ccc} \sigma^a/2 &0 & 0 \\
0 & 0 & 0\\ 0 & 0 & 0
\end{array}\right), \ \ \ &Y_1=
{\rm diag}(3,3,-2,-2,-2)/10\nonumber \\
&Q_2^a=\left( \begin{array}{ccc} 0 & 0 & 0\\
0 & 0 & 0 \\
0 &0&-\sigma^{a*}/2\end{array} \right), & Y_2={\rm
diag}(2,2,2,-3,-3)/10~,
\end{eqnarray}
where $\sigma^a$ are the Pauli $\sigma$ matrices. The $Q^a$'s are
$5 \times 5$ matrices written in terms of $2 \times 2$, 1, and $2
\times 2$ blocks.  The vacuum breaks the $[SU(2)\times U(1)]^2$
gauge symmetry down to the diagonal subgroup, giving one set of
$[SU(2) \times U(1)]$ gauge bosons masses of order $f$, while the
other set are left massless, and are identified as the $SU(2)_L \times U(1)_Y$
gauge fields of the standard model. The Goldstone boson matrix
$\Pi$ is given by
\begin{equation}
\Pi = \left( \begin{array}{ccc}
0 & \frac{H}{\sqrt{2}} & \Phi
\\ \frac{H^\dagger}{\sqrt{2}}& 0 & \frac{H^T}{\sqrt{2}}\\ \Phi^\dagger
&\frac{H^*}{\sqrt{2}} & 0
\end{array}\right),
\end{equation}
where $H$ is the little Higgs doublet $(h^+,h)^T$ and $\Phi$
is a complex triplet under $SU(2)_L$ which forms a symmetric tensor $\Phi_{ij}$
with components $\phi^{++},\phi^+,\phi^0$ and a pseudoscalar
$\phi^P$ as defined in~\cite{han1}.  $\phi^0$ and $\phi^P$ are both
real scalars.  The Goldstone bosons which
are eaten to become the longitudinal modes of the partners of the
standard model gauge fields are set to zero in the pion matrix,
as we have gone to unitary gauge.

The underlying idea for implementing T-parity is to assign all
non-SM particles odd parity, and all SM particles even parity,
thus avoiding tree level constraints from EWPT. It turns out that
there will be one case in which this assignment is not possible
when we discuss the top quark sector of the model, however for the
gauge and scalar sectors all non-SM particles are T-odd.  To
implement T-parity in the gauge sector one notices that the
$\Sigma_0$ VEV separates the gauged generators into a broken set
$\{Q_1^a-Q_2^a,Y_1-Y_2\}$ and an unbroken set
$\{Q_1^a+Q_2^a,Y_1+Y_2\}$ of generators. Using the action of the
$\mathrm{Z}_2$ inner automorphism discussed earlier,
$T^a\rightarrow T^a$ for unbroken generators, and $X^a\rightarrow
-X^a$ for broken generators, a natural action of T-parity on the
gauge fields is defined as
\begin{equation}
\label{tpargauge}
 A_1 \leftrightarrow A_2,
\end{equation}
where $A_1,A_2$ are the gauge fields corresponding to the
$[SU(2)\times U(1)]_1$ and $[SU(2)\times U(1)]_2$ gauge groups,
respectively.  The action for T-parity in the scalar sector is
defined as
\begin{equation}
\label{tparscalar}
\Pi \to - \Omega \Pi \Omega,
\end{equation}
where $\Omega = \mathrm{diag} (1,1,-1,1,1)$.   $\Omega$ is
introduced to give the Higgs positive parity while keeping
the triplet odd.  From the transformation of $\Pi$ we can also
write the transformation law for $\Sigma$ under T-parity
\begin{equation}
\Sigma\rightarrow \Sigma_0 \Omega \Sigma^\dagger
\Omega \Sigma_0 \equiv \tilde{\Sigma}.
\end{equation}

The kinetic term of the non-linear $\sigma$ model field $\Sigma$ in the
littlest Higgs is
\begin{equation}
\label{eq:sigkin} \frac{f^2}{8} {\rm Tr} D_\mu \Sigma
(D^\mu\Sigma)^\dagger,
\end{equation}
where
\begin{equation} D_\mu
\Sigma =
\partial_\mu \Sigma - i \sum_j \left[ g_j W_j^a (Q_j^a \Sigma +
\Sigma Q_j^{aT} ) + g'_j B_j( Y_j \Sigma + \Sigma Y_j)\right],
\end{equation}
with $j=1,2$.  The kinetic terms for the sigma field and gauge
bosons are invariant under T-parity as written down originally in
the littlest Higgs model with the additional proviso that
$g_1=g_2=\sqrt{2} g$ and $g_1'=g_2'=\sqrt{2}g'$.  Since T-parity
exchanges the gauge fields (\ref{tpargauge}), the gauge couplings must
be equal for the Lagrangian to be invariant.  We identify $g$ and $g'$
with the SM $SU(2)$ and $U(1)_Y$ gauge couplings respectively.

In the gauge sector before EWSB there is a linear combination of
gauge bosons that acquire a mass of order $f$ from
(\ref{eq:sigkin}),
\begin{equation}
\begin{array}{cc}
W_H^a = \frac{1}{\sqrt{2}}(W_1^a - W_2^a), & M_{W^a_H} = g f, \\
B_H= \frac{1}{\sqrt{2}}(B_1-B_2), & M_{B_H} =
\frac{g'}{\sqrt{5}}f,
\end{array}
\end{equation}
while the linear combinations
\begin{equation}
\begin{array}{cc}
W_L^a = \frac{1}{\sqrt{2}}(W_1^a + W_2^a), & B_L =
\frac{1}{\sqrt{2}}(B_1+B_2),
\end{array}
\end{equation}
remain massless and are identified with the SM gauge bosons. From
the T-parity transformation (\ref{tpargauge}) the heavy gauge
bosons are odd under T-parity while the SM gauge bosons are even.
After EWSB the VEV of the Higgs, $\langle
H\rangle^T=(0,v/\sqrt{2})$, will shift the mass eigenstates in the
heavy gauge boson sector. The new mass eigenstates in the neutral
heavy sector will be a linear combination of the $W^3_H$ and the
$B_H$ gauge bosons, producing an $A_H$ and a $Z_H$.  The mixing
angle introduced into the neutral heavy sector by EWSB will be of
order $v^2/f^2$:
\begin{equation}
\sin \theta_H \approx \frac{5 g g'}{4 (5 g^2 - g'^2)}
\frac{v^2}{f^2}.
\end{equation}
The new heavy neutral mass eigenstates are given by
\begin{equation}
\begin{array}{cc}
Z_H = \sin \theta_H B_H + \cos \theta_H W^3_H, &  M^2_{Z_H}=g^2 f^2-\frac{g^2 v^2}{4}\\
A_H = \cos \theta_H B_H - \sin \theta_H W^3_H, &
 M^2_{A_H}=\frac{g'^2 f^2}{5}-\frac{g'^2 v^2}{4}\label{heavyphot}
\end{array}
\end{equation}
while a set of heavy charged gauge bosons can be written as
\begin{equation}
\begin{array}{cc}
W_H^{\pm}=\frac{1}{\sqrt{2}}(W_H^1\mp i W_H^2),  &
M^2_{W_H^{\pm}}=g^2 f^2-\frac{g^2 v^2}{4}.
\end{array}
\end{equation}

In the scalar sector from the T-parity transformation
(\ref{tparscalar}) one can see that the Higgs doublet has positive
parity under this transformation, and that the $SU(2)_L$ triplet
$\Phi$ has odd parity.  We will briefly recall that in a little
Higgs model, EWSB is generated radiatively through the
Coleman-Weinberg potential.  The difference from~\cite{littlest} in
this sector with the addition of T-parity is
that the coupling $H \Phi H$ is forbidden by
T-parity.  The absence of the $H \Phi H$ coupling forbids a dangerous
non-zero triplet VEV\cite{us1,us2} after EWSB.   The
mass of $\Phi$ can be related to the mass of the Higgs from the
Coleman-Weinberg potential~\cite{han1} to give
\begin{equation}\label{tripletmass}
m^2_{\Phi}=\frac{2 m_H^2 f^2}{v^2},
\end{equation}
where all components of the triplet are degenerate at the order we
are examining.

It summary, the implementation of T-parity for the gauge sector
has essentially decoupled the light and heavy sectors by
setting $g_1=g_2$ and $g'_1=g'_2$. T-parity has also forbidden the
VEV of the triplet $\Phi$.   Most constraints from EWPT on the
original littlest Higgs~\cite{us1,slac,us2} vanish in this limit.

To fully implement T-parity in a consistent way we must now turn
our attention to the fermion sector.  We would like to introduce
SM fermions that transform linearly under the gauge symmetries to
avoid large contributions to four fermion operators that would
require the scale $f$ to be large~\cite{tpar3}.  Since T-parity
exchanges $SU(2)_1$ and $SU(2)_2$ one must introduce two doublets
$\psi_1$, and $\psi_2$, which transform linearly under $SU(2)_1$
and $SU(2)_2$ respectively.  These doublets are mapped into each
other under the action of T-parity.  To obtain the SM at low
energy we would like to give an $f$ scale mass to the T-odd linear
combination of these doublets.  In giving mass to only the T-odd
linear combination,  at this stage one should introduce a
``mirror" fermion with odd parity and write down a mass term with
the T-odd linear combination of $\psi_1$ and
$\psi_2$~\cite{tpar3}. Since T-parity exchanges the two $SU(2)$'s,
to avoid introducing an additional copy of the mirror fermion, it
must have a non-linear symmetry transformation.

We will now explain in detail how to make a gauge invariant
term that only gives mass to the T-odd linear combination of
$\psi_1$ and $\psi_2$.  For those readers only interested in the
phenomenology it is not crucial to understand where in the end
the expression (\ref{heavyyuk}) comes from, but it is important to
note that there will be additional interactions apart from just mass
terms in (\ref{heavyyuk}).  A set of mirror fermions $\Psi'$ is
introduced in a complete multiplet of $SO(5)$, whose
transformation under $SU(5)$ is non-linear. The mirror fermions
must be introduced in a complete multiplet in order to eliminate
potential quartic divergences in the Higgs mass~\cite{tpar2}.  In
order to give this $\Psi'$ multiplet interactions with other
fields which obey linear transformation laws, we introduce a field
$\xi = e^{i \Pi/f}$. This technology, for those readers familiar
with it, is reminiscent of introducing baryons in a nonlinear
chiral $SU(3)_L \times SU(3)_R$ model~\cite{georgi}. In terms of
$\xi$, the field $\Sigma$ can be expressed as $\Sigma = \xi^2
\Sigma_0$.  From the linear transformation of $\Sigma$, we infer that the field $\xi$ has the following
transformation under a global $SU(5)$ rotation $V$:
\begin{equation}
\Sigma \rightarrow V \Sigma V^T\ \ \Rightarrow\ \ \xi \rightarrow U \xi \Sigma_0 V^T \Sigma_0 = V \xi U^\dagger
\end{equation}
where $U$ takes values in the Lie algebra of the unbroken $SO(5)$
subgroup, and is a function of both $V$ and the pion fields. It is
this same $U$ under which the $\Psi'$ multiplet transforms:
\begin{equation}
\Psi' \rightarrow U \Psi'.
\end{equation}
The fermion doublets $\psi_1,\psi_2$ can be embedded into
incomplete representations $\Psi_1,\Psi_2$ of $SU(5)$, and the
field content can be expressed as follows:
\begin{equation}
\begin{array}{ccc}
\Psi_1=\left(\begin{array}{c} \psi_1 \\ 0 \\ 0 \end{array}\right)
& \Psi_2=\left(\begin{array}{c} 0 \\ 0 \\ \psi_2
\end{array}\right) &
\Psi'=\left(\begin{array}{c} \tilde{\psi}' \\ \chi' \\ \psi'
\end{array}\right)
\end{array}
\end{equation}
where $\chi'$ is a singlet and $\tilde{\psi}'$ is a doublet
under $SU(2)_2$.
The transformation laws for $\Psi_1$ and $\Psi_2$ are as follows:
\begin{equation}
\Psi_1 \rightarrow V^* \Psi_1 \hspace{.2in} \Psi_2 \rightarrow V
\Psi_2
\end{equation}
The action of T-parity on the multiplets takes
$\Psi_1\leftrightarrow -\Sigma_0 \Psi_2$ and $\Psi' \rightarrow
-\Psi'$. One can now write down a Yukawa-type interaction to give
masses to the mirror fermions,
\begin{equation}\label{heavyyuk}
\kappa f (\bar{\Psi}_2 \xi \Psi'+\bar{\Psi}_1 \Sigma_0 \Omega
\xi^\dagger \Omega \Psi'),
\end{equation}
which is invariant under a global $SU(5)$ rotation, and also under
the action of T-parity.  From (\ref{heavyyuk}) one fermion doublet
$\psi_H=\frac{1}{\sqrt{2}}(\psi_1+\psi_2)$ acquires a mass $\kappa
f$, while the other combination
$\psi_{SM}=\frac{1}{\sqrt{2}}(\psi_1-\psi_2)$ remains massless and
is identified with the SM doublets.  To give masses to the
remaining fields $\chi'$ and $\tilde{\psi}'$ one must introduce
additional fermions $\chi$ and $\tilde{\psi}$ along with another
singlet $\tilde{\chi}$ which occupy a spinor representation of
$SO(5)$, $S$, and then introduce Dirac mass terms for these fields.

After introducing the Yukawa coupling (\ref{heavyyuk}) which gives
the mirror fermions mass we need to introduce kinetic terms for
the fermions in this model.  Since $\Psi_1$ and $\Psi_2$ transform
linearly, their kinetic terms are straightforward to write down:
\begin{equation}\label{lkinsm}
\mathcal{L}_{kin} \supset \bar{\Psi}_1 \bar{\sigma}^{\mu}
D^1_{\mu} \Psi_1+\bar{\Psi}_2 \bar{\sigma}^{\mu} D^2_{\mu} \Psi_2,
\end{equation}
where
\begin{eqnarray}
D^1_{\mu} &=&\partial_{\mu}- i \sqrt{2} g Q_1^a W_1^a-i \sqrt{2} g'
Y_1^{(\Psi_1)} B_{1\mu} - i \sqrt{2} g' Y_2^{(\Psi_1)} B_{2\mu}, \\
D^2_{\mu} &=&\partial_{\mu} + i \sqrt{2} g (Q_2^a)^T W_1^a - i
\sqrt{2} g' Y_1^{(\Psi_2)} B_{1\mu} - i \sqrt{2} g' Y_2^{(\Psi_2)}
B_{2\mu},
\end{eqnarray}
 and the $U(1)$ charges $Y_i^{(\Psi_j)}$ are given in Table
\ref{tb:charges} which will can be determined from gauge
invariance of the Yukawa couplings and T-parity. Re-expressing
(\ref{lkinsm}) in terms of the mass eigenstates found from
(\ref{heavyyuk}) one obtains
\begin{equation}\label{lkinsmmass}
\mathcal{L}_{kin} \supset \bar{\psi}_{SM} \bar{\sigma}^{\mu}
D^L_{\mu} \psi_{SM}+\bar{\psi}_{H} \bar{\sigma}^{\mu} D^L_{\mu}
\psi_{H},
\end{equation}
which contains the usual kinetic terms for the SM doublet fermions
where $D^L_{\mu}$ is the usual SM covariant derivative, as well as
kinetic terms for the heavy fermions.  In addition to the standard
kinetic terms in~(\ref{lkinsmmass}) there are interaction terms of
the form
\begin{equation}\label{kininteract}
c \bar{\psi}_{SM} \bar{\sigma}^\mu V_{H\mu} \psi_H,
\end{equation}
where $c$ is a gauge coupling and $V_H$ is a heavy gauge boson.
These new interactions between heavy and light fields come from
expressing~(\ref{lkinsm}) in mass eigenstates.  To write down an
invariant kinetic term for the $\Psi'$ field we must make use of
the $\xi$ field in the following way
\begin{equation}\label{kinmirror}
\bar{\Psi}' \bar{\sigma}^\mu (\partial_\mu+\xi^\dagger D_\mu \xi
+\xi D_\mu^T \xi^\dagger) \Psi',
\end{equation}
where
\begin{eqnarray}\label{mirrorcov}
D_\mu &=& \partial_\mu - i c V_L Q_V - i c V_H Q_A, \arline
D_\mu^T &=& \partial_\mu - i c V_L Q_V + i c V_H Q_A.
\end{eqnarray}
$D_\mu^T$ is the T-parity transformed covariant derivative.
One can write down a similar kinetic term for the spinor field $S$
by expressing the covariant derivatives in Eq.~(\ref{mirrorcov}) in the spinor
representation of $SO(5)$.  There are dangerous four fermion operators
arising at the one loop level which the T-odd partners of the standard
model left handed doublets serve to cut off, thus these fermions
cannot be taken to be arbitrarily heavy.  We discuss this in more
detail in Appendix A.  For the purposes of studying a simpler, and  model
independent phenomenology, we neglect these fermions in this paper,
leaving their study for future work.

In order to avoid dangerous contributions to the Higgs mass from
one loop quadratic divergences, the third generation Yukawa sector
must be modified so that it incorporates the collective symmetry
breaking pattern of \cite{littlest}.  In order to do this, the $\Psi_1$
and $\Psi_2$
multiplets for the third generation must be completed to
representations of the $SU(3)_1$ and $SU(3)_2$ subgroups of the
full $SU(5)$.  These are
\begin{equation}
\begin{array}{ccc}
Q_1=\left(\begin{array}{c} q_1 \\ t'_1 \\ 0 \end{array}\right)
& Q_2=\left(\begin{array}{c} 0 \\ t'_2 \\ q_2
\end{array}\right),
\end{array}
\end{equation}
where $Q_1$ and $Q_2$ obey the same transformation laws under T-parity
and the $SU(5)$ symmetry as
do $\Psi_1$ and $\Psi_2$.  It should be noted that the quark doublets
are embedded such that
\begin{equation}
q_i = -i \sigma_2 \left(\begin{array}{c} t_i \\ b_i \end{array}\right).
\end{equation}
One must also introduce additional singlets $t'_{1R}$ and
$t'_{2R}$ which transform under T-parity as
\begin{equation}
t'_{1R}\leftrightarrow -t'_{2R}
\end{equation}
so the top sector masses can be generated in the following
T-parity invariant way
\begin{eqnarray}\label{topyuk}
\mathcal{L}_t &=& \frac{1}{2 \sqrt{2}}\lambda_1 f \epsilon_{ijk}
\epsilon_{xy} \big[ (\bar{Q}_1)_i(\Sigma)_{jx} (\Sigma)_{ky} -
(\bar{Q}_2 \Sigma_0)_i (\tilde{\Sigma})_{jx} (\tilde{\Sigma})_{ky}
\big] u_{3R} \arline && \hspace{1in} + \lambda_2 f (\bar{t}'_1
t'_{1R} + \bar{t}'_2 t'_{2R})+ h.c.
\end{eqnarray}
We point out that this is a slightly different implementation of
the top Yukawa coupling than in~\cite{tpar3}; T-parity and gauge invariance
require the couplings of $t'_1$ and $t'_2$ to be equal and leads
to new phenomenology for the heavy top quarks. This Yukawa term
generates a mass for the top quark given by
\begin{equation}
m_{\mathrm{top}} =\frac{ \lambda_1 \lambda_2 v}{\sqrt{\lambda_1^2+\lambda_2^2}},
\end{equation}
while the T-even combination of $t'_1$ and $t'_2$ has mass
\begin{equation}\label{tevenmass}
m_{t'_+} = \sqrt{\lambda_1^2+\lambda_2^2} f,
\end{equation}
and the T-odd combination has mass
\begin{equation}\label{toddmass}
m_{t'_-} = \lambda_2 f,
\end{equation}
with $t'_{\pm} = \frac{1}{\sqrt{2}} (t'_1 \mp t'_2)$.  The T-odd
combination of the $q_1$ and $q_2$ doublets obtains a mass through
a Yukawa identical to Equation (\ref{heavyyuk}).  The other two
generations of up-type quarks acquire their mass through similar
terms, though with the $t'$ quarks missing from the $Q_1$ and
$Q_2$ multiplets since the Yukawa couplings are small and one does
not have to worry about quadratic divergences.  Requiring that the
top-sector Yukawa term be gauge invariant determines the $U(1)$ charges
of the fermions up to one degree of freedom, which is then fixed
by imposing T-parity.  The resulting charges are given in Table
\ref{tb:charges}.

\begin{table}[h]
\center{\begin{tabular}{|c|l||c|l|} \hline $q_1$ & $(\mathbf{2}
,1/30 ; \mathbf{1} , 2/15 )$ & $q_2$ & $(\mathbf{1}
, 2/15 ; \mathbf{2} , 1/30 )$ \\
\hline $t'_1$ & $(\mathbf{1} , 8/15 ; \mathbf{1} , 2/15 )$ &
$t'_2$ & $(\mathbf{1}
, 2/15 ; \mathbf{1} , 8/15 )$ \\
\hline $t'_{1R}$ & $(\mathbf{1} , 8/15 ; \mathbf{1} , 2/15 )$ &
$t'_{2R}$ & $(\mathbf{1}
, 2/15 ; \mathbf{1} , 8/15 )$ \\

\hline $u_{3R}$ & $(\mathbf{1} , 1/3 ; \mathbf{1} , 1/3 )$ & $d_R$
&
$(\mathbf{1}, -1/6, \mathbf{1},-1/6)$  \\
\hline $l_1$ & $(\mathbf{2} , -1/5 ; \mathbf{1} , -3/10 )$ & $l_2$
&
$(\mathbf{1}, -3/10, \mathbf{1},-1/5)$ \\
\hline $e_R$ & $(\mathbf{1} , -1/2 ; \mathbf{1} , -1/2 )$ & & \\
\hline
\end{tabular}}
\caption{The $\left[ SU(2)_1 \times U(1)_1 \right] \times
  \left[ SU(2)_2 \times U(1)_2 \right]$ quantum numbers of the fermion
  fields that are required to make Eqns. (\ref{heavyyuk}),(\ref{topyuk}), and (\ref{downyuk}) gauge invariant.
\label{tb:charges} }
\end{table}

We also need to construct a Yukawa interaction which gives the
down-type quarks a mass after electroweak symmetry breaking.  The
following term accomplishes this:
\begin{equation}\label{downyuk}
\mathcal{L}_d = \frac{1}{4}\lambda_d f \epsilon_{ij} \epsilon_{xy}
\big[ (\bar{\Psi}'_2)_x (\Sigma)_{3i} (\Sigma)_{jy} -
(\bar{\Psi}'_1 \Sigma_0)_x (\tilde{\Sigma})_{3i}
(\tilde{\Sigma})_{jy}  \big] d_R + h.c.
\end{equation}
The doublets are embedded in $\Psi'_1$ and $\Psi'_2$ such that
\begin{equation}
\begin{array}{cc}
\Psi'_1 = \left(\begin{array}{c} -i \sigma_2 q_1 \\ 0 \\ 0 \end{array}\right)
& \Psi'_2=\left(\begin{array}{c} 0 \\ 0 \\ -i \sigma_2 q_2
\end{array}\right),
\end{array}
\end{equation}
and they transform under $SU(5)$ as $\Psi'_1 \rightarrow V
\Psi'_1$, and $\Psi'_2 \rightarrow V^* \Psi'_2$.  The lepton
Yukawas can be taken to be identical to Equation (\ref{downyuk}).

\begin{figure}[t]
\centerline{\includegraphics[width=0.6\hsize]{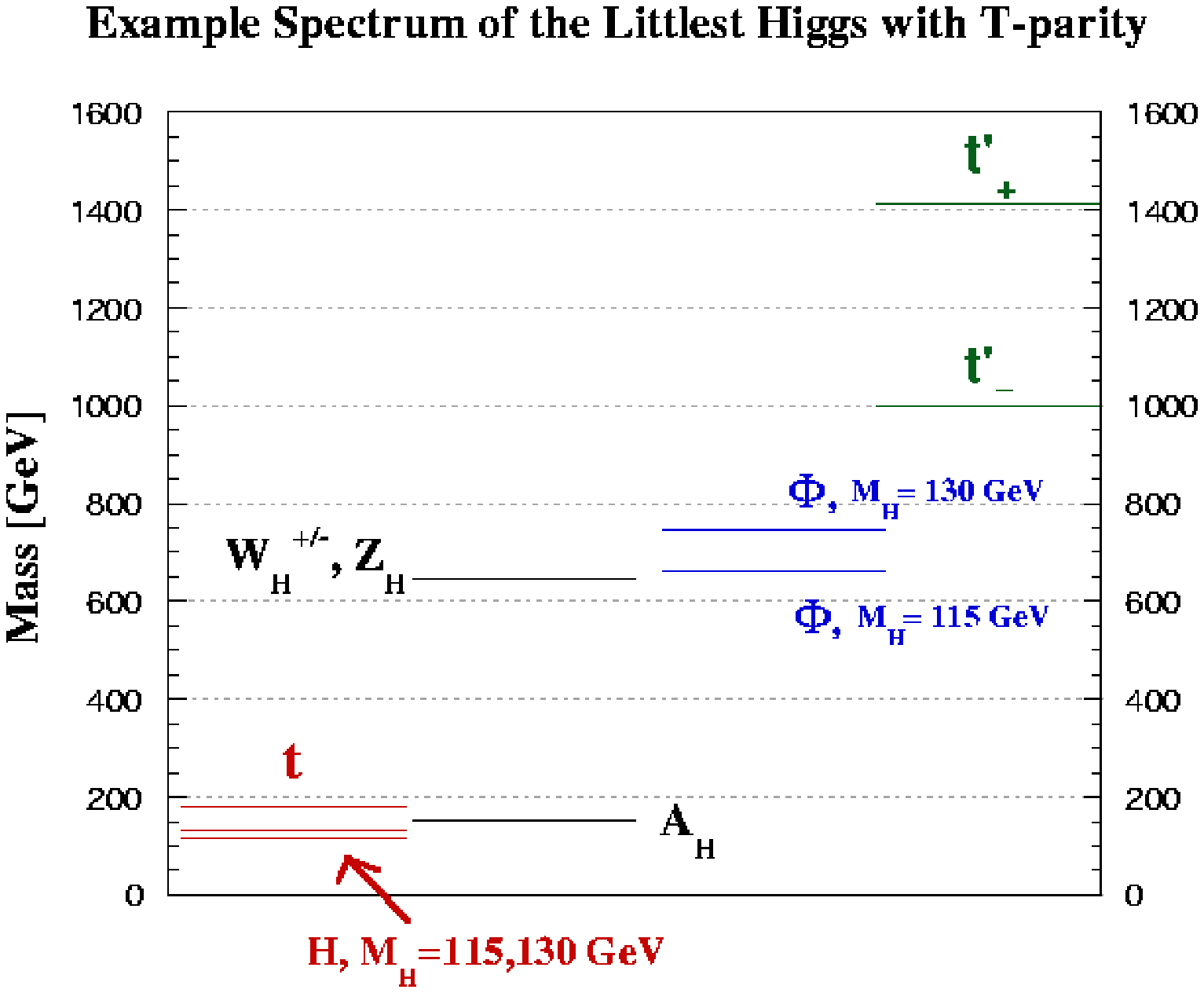}}
\caption{We plot a sample spectrum for the littlest Higgs with
T-parity.  The top quark mass and two values of the Higgs mass are
plotted as a reference.  The spectrum of heavy particles is
plotted for $f=1$~TeV.  The $\Phi$ mass is plotted for two
different values of the Higgs mass, $M_H=115,130$~GeV. A value of
$s_\lambda=\frac{1}{\sqrt{2}}$ is used to determine the masses of
$t'_+$ and $t'_-$.}\label{spectra}
\end{figure}

The final spectrum that we consider has some important features that
we briefly summarize.  In Figure \ref{spectra} we give an example
spectrum for a typical choice of model parameters.
The lightest new particle introduced in this model
is the $A_H$.  The array of scalars coming from the $SU(2)_L$
triplet can be either lighter than, or more massive than the $W_H$
and $Z_H$ gauge bosons, depending on the values of $f$ and $m_H$.  For
smaller $m_H$ and $f$ smaller than
$1$~TeV, the scalar triplet is lighter than these gauge fields,
but for larger $f$ and $m_H$ the triplet is in fact more massive.
A useful parameter that we will use to discuss the new fermions, $\tp$
and $\tm$, is
\begin{equation}
s_\lambda \equiv \frac{\lambda_2}{\sqrt{\lambda_1^2 +
\lambda_2^2}} = \frac{m_{\tm}}{m_{\tp}}.
\end{equation}
One thing to notice in particular is that the $\tm$ fermion is
always lighter than the $\tp$.  This will lead to novel phenomenology
for the $\tp$ that was not present before the introduction of
T-parity.

\subsection{Alternative implementations of
T-parity}\label{alternatives}

We will now turn our discussion to other implementations of
T-parity~\cite{tpar3} in the littlest Higgs, to give credence to
why we will study the phenomenology of only the model described in
Section~\ref{subsec:tparity} in detail.  In the original 3-site
moose model of T-parity described in~\cite{tpar1}, three copies of
$SU(2)\times U(1)$ were gauged, one at each site.  In this model,
T-parity was realized as a symmetry exchanging two of the sites,
while the third site was neutral under the action of T-parity.
This kind of set up enabled fermions to be introduced in linear
representations which naturally avoids the constraints imposed
when SM fermions have non-linear gauge
transformations~\cite{tpar3}.  One can come up with an analogous
set up to the three-site moose model with a littlest Higgs
structure~\cite{tpar3} by enlarging the global symmetry of the
littlest Higgs and gauging a third copy of $SU(2)\times U(1)$.
In~\cite{tpar3} the specific implementations of this idea were
accomplished by enlarging the global structure $SU(5)/SO(5)$ to
$(SU(5)\times G_r)/SO(5)$ where $G_r$ was either $SO(5)$ or
$SU(5)$.  The same generators as in~(\ref{gauged}) are gauged
along with an additional $SU(2)\times U(1)$
\begin{eqnarray}
&Q_r^a=\left( \begin{array}{ccc} \sigma^a/2 &0 & 0 \\
0 & 0 & 0\\ 0 & 0 & -\sigma^{a*}/2
\end{array}\right), \ \ \ &Y_r=
{\rm diag}(1,1,0,-1,-1)/2.
\end{eqnarray}
One should immediately note that in this class of models with a
third set of gauge groups which is neutral under T-parity, there
will be new T-even heavy gauge bosons.  In addition, there will
also be T-even heavy scalars. Having T-even heavy gauge bosons and
scalars is a rather dangerous prospect given that EWPT normally
require the scale $f$ to be large, which reintroduces a fine
tuning~\cite{us1,us2}.  In this class of model with three sets of
$SU(2)\times U(1)$ gauge groups the problem of raising the scale
$f$ is avoided by taking the gauge couplings $g_r$ of the third
set to be $g_r\sim 4\pi$, which decouples the heavy T-even gauge
bosons.  The new heavy T-even scalars can be decoupled by
introducing strongly coupled plaquette operators which raise the
scalar masses to $\mathcal{O}(10\, \mathrm{TeV})$.

If one is willing to take the limits just described to
avoid the constraints of EWPT, the resulting theory will be
identical in the gauge and scalar sectors (around the TeV scale
and below) to the littlest Higgs with T-parity discussed in
Section~\ref{subsec:tparity}.  The only difference between the
implementation discussed in Section~\ref{subsec:tparity} and the
$(SU(5)\times G_r)/SO(5)$ models arises in the fermion sector. All
of the implementations of the fermion sector have a mirror fermion
type implementation of the SM fermions (e.g. (\ref{heavyyuk})). The
type of $\Sigma$ or $\xi$ fields available in the particular
model~\cite{tpar3} will dictate the number of fermions required to
write down a heavy Yukawa to lift the mass of one set of fermions
as in (\ref{heavyyuk}).  One thing that must be pointed out is
that in the top sector, all three types of littlest Higgs with
T-parity have the same spectrum.  Each model contains a T-odd and
T-even partner of the top quark at the TeV scale where all
implementations share the same
couplings (\ref{topyuk}). In the heavy Yukawa sector as mentioned before, the
number of fermions is different in the three implementations but
it is important to point out that the masses of the new fermions
are constrained.  In the little Higgs models with T-parity,
generically one requires some of the mirror fermion doublets to have
$\mathcal{O}(\mathrm{TeV})$ masses or lower to avoid constraints on four
fermion operators.

In analyzing the phenomenology of the littlest Higgs with
T-parity we use the model outlined in Section~\ref{subsec:tparity}
for the following reason.  Since the CCWZ implementation has an
identical scalar and gauge sector at around a TeV (because of the
EW constraints on the two other implementations), one does not have
to rely on strong coupling and one captures much of the interesting
phenomenology that does not rely on the details of the fermion
spectrum.   Since the top Yukawa sector of the model is
identical in all three implementations we will capture its
phenomenology in any of the models.  In this paper we will choose to
ignore all T-odd heavy fermions except for the $t'_-$.
Further analysis of the fermion sector would be an interesting
future project to see the implications of more T-odd
fermions at the TeV scale, both in collider studies, and in
electroweak precision.

\section{The $A_H$ as a dark matter candidate}
\label{darkmatter}
If we require that T-parity be an exact, or nearly exact,
symmetry, the lightest new particle introduced is stable.  While
EWP does not require such a large suppression of the tree level
operators, it is interesting to pose the question of whether
T-parity is part of a more fundamental symmetry arising from the
ultraviolet completion of the little Higgs mechanism.  If this
particle is also neutral, it provides a promising candidate for
WIMP dark matter.  This particle will be in equilibrium with the
thermal bath at early times in the history of the universe, being
pair produced in collisions of lighter standard model particles,
and annihilating via the same channels.  As the universe cools
however, these processes will fall out of equilibrium, and the
number density of the lightest parity odd particle (LPOP) will
begin to decrease. This happens until the rate of expansion of the
universe overtakes the annihilation rate, at which point the
abundance of the LPOP will freeze out.  From this point on the
relic abundance of the LPOP will simply follow the expansion rate.
This process is described by a Boltzmann equation
\begin{equation}\label{boltz}
\frac{d n_\chi}{dt} + 3 H n_\chi = -\langle \sigma_A v \rangle
\left[ n_\chi^2 -
  (n_\chi)_{\mathrm{eq}}^2 \right],
\end{equation}
where $\langle \sigma_A v \rangle$ is the thermally averaged
annihilation coefficient for the WIMP $\chi$.  $H$ is the Hubble
constant, $n_\chi$ is the number density of the WIMP in the
thermal bath, and $(n_\chi)_\mathrm{eq}$ is the equilibrium number
density.  The annihilation cross section is given by
\begin{equation}
\sigma_A = \sigma \left( 2 \chi \rightarrow 2 X \right).
\end{equation}
The parameters that describe the interactions and mass of the WIMP
are all contained in the annihilation cross section.

Unlike other types of little Higgs models that have scalar dark
matter candidates ~\cite{andreasjay}, in the model that we have
described, the LPOP is the $A_H$, the little Higgs partner of the
standard model hypercharge gauge boson, which we refer to as a
heavy photon (see eq (\ref{heavyphot})).  As shown in Section
\ref{sec:model}, the mass of this field to zeroth order in the
expansion about the electroweak vacuum is directly related to the
breaking scale of the $SU(5)$ global symmetry, $f$.
\begin{equation}
M_{A_H} = \frac{g' f}{\sqrt{5}},
\end{equation}
where $g'$ is the standard model hypercharge gauge coupling.  The
partner of the photon is the LPOP due to the small gauge coupling,
$g'$, as well as the factor of $\sqrt{5}$ that comes from the
$SU(5)$ normalization of the $U(1)$ generators.  The couplings of
the $A_H$ to the mass eigenstates are determined by the breaking
scale $f$, and parameters in the Yukawa sector. However, the rate
of annihilation to top quarks is small, therefore the annihilation
coefficient is only weakly dependent on the new Yukawa couplings.
The only remaining degree of freedom which governs the
annihilation cross section is the mass of the Higgs. We calculate
the annihilation cross section using COMPHEP \cite{comphep}, and
based on the late time solution to (\ref{boltz}), evaluate the
relic density of the $A_H$ dark matter.  We use cosmological
constraints on relic abundance of dark matter to find constraints
on $f$.  Of particular interest is the fact that relic abundance
considerations put upper bounds on the breaking scale $f$, rather
than just lower bounds.  One could also use dark matter search
constraints to find additional bounds on $f$, however we leave
such work for future study.

\subsection{Relic abundance calculation}

The annihilation rate at a given center of mass velocity can be
expanded in the non-relativistic limit as follows:
\begin{equation}
\sigma_A v = a + b v^2 + \cdots.
\end{equation}

The coefficients $a$ and $b$ are determined by the couplings of
the model being studied.  If $a$ is non zero, the dark matter
candidate is referred to as an `s-annihilator', and a
`p-annihilator' otherwise \cite{maximkonst}.  It is important here
to distinguish the interactions of the little Higgs dark matter
from SUSY dark matter. In supersymmetric theories, the WIMP is
generally a p-annihilator, whereas in little Higgs theories, it
will be an s-annihilator.  This generally produces larger
annihilation cross sections, and thus lower relic abundances.

In some models that contain dark matter, there are other particles
which have a mass that is very close to that of the WIMP.  In such
cases, the relic abundance may be additionally depleted by
processes that convert the WIMP to this nearly degenerate
particle, which is then able then able to itself annihilate with
other particles in the thermal bath.  This process is called
coannihilation.  In the model that we consider, the $A_H$ is
always much lighter than all of the other T-odd particles, as can
be seen in the following formula:
\begin{equation}
\frac{m_{A_H}}{m_\phi} \approx \frac{g' v}{\sqrt{10} m_H} \approx
.24 \ \ \ \frac{m_{A_H}}{m_{Z_H}} \approx \frac{g'}{\sqrt{5} g}
\approx .24,
\end{equation}
where we have taken $m_H = 115$ GeV.  Therefore coannihilation is
not relevant in this model for the spectrum considered.

The techniques for solving the Boltzman equation for the
population of the WIMP are well studied.  One can numerically
solve this equation, or one can use an approximate analytic
result.  This is generated by solving the early and late time
behavior of the differential equation, and then matching in an
intermediate region. We do not repeat this calculation, but simply
quote the result.  The relic abundance is given approximately by
\begin{equation}
\Omega_\mathrm{dm} h^2 = \frac{1.07 \times 10^9 \mathrm{GeV}^{-1}
  M}{g_*^{1/2} m_{\mathrm{Pl}} \langle \sigma_A v \rangle T_F},
\end{equation}
where $M$ is the mass of the WIMP, and $T_F$ is the freezeout
temperature, the temperature at which the Hubble term begins to
dominate over the annihilation term. The term $g_*$ is related to
the number of relativistic degrees of freedom that exist at the
time of freezeout.  The freezeout temperature is dependent on the
mass of the WIMP, though only weakly, and is determined through
the solution of (\ref{boltz}). For most WIMP dark matter, the
freezeout temperature is roughly $M/20$.  In this model, the
number that we calculate is $T_F \approx M_{A_H}/22.5$ for most
ranges of $f$ and $m_H$.

\begin{figure}[t]
\centerline{\includegraphics[width=0.3\hsize]{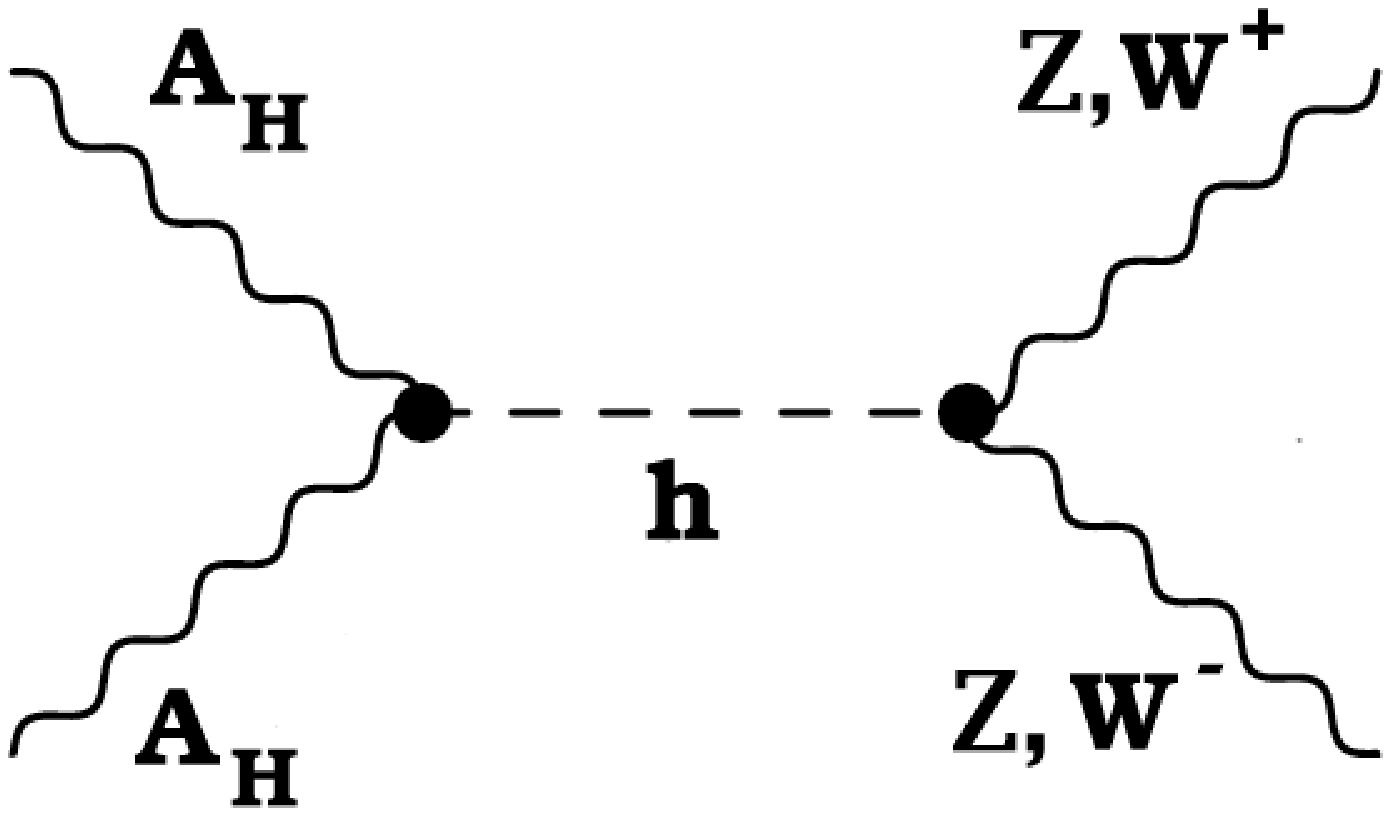}\includegraphics[width=0.6\hsize]{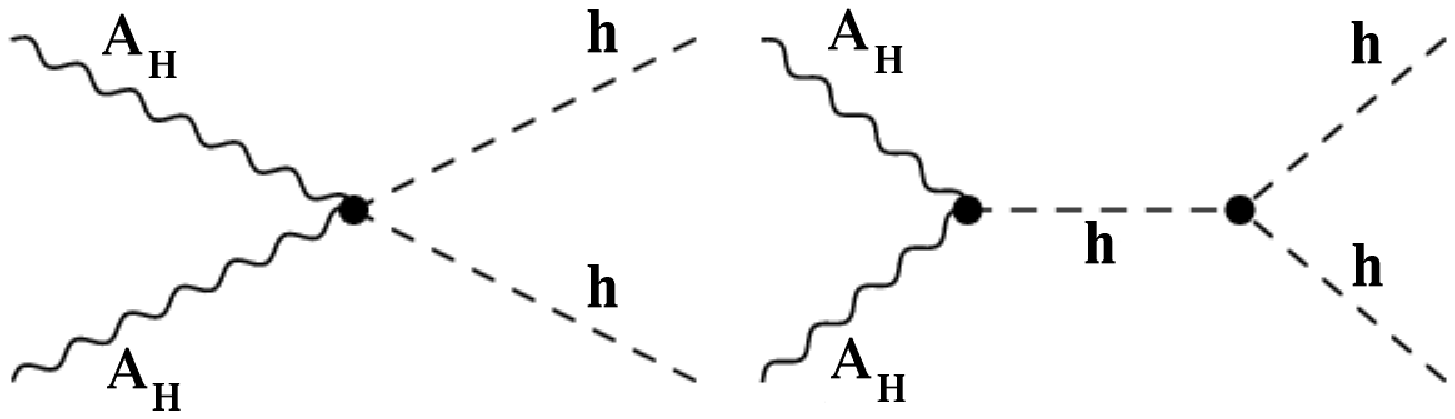}}
\caption{The $A_H$ annhilates predominantly to SM gauge and Higgs
  bosons.  These are the diagrams which give the largest contributions
  to the annihilation coefficient $\langle \sigma_A v \rangle$ for the
  ranges of $f$ and $m_H$ that we examine.}
\label{annchan}
\end{figure}

If the cross section is too small, the WIMP does not annihilate
enough before freeze out, and the relic density is larger than the
experimental bounds allow. A somewhat less disturbing issue arises
when the cross section is too large, and the relic density is too
small to account for the total fraction of dark matter.  In this
situation, there must be another dark matter candidate to account
for the remaining fraction, such as axions, which does not arise
as part of the T-parity little Higgs mechanism.  Current data
coming from numerous astrophysical observables, most notably the
WMAP sky survey~\cite{wmap}, place the dark matter density (in
units of the critical density) at
\begin{equation}
\Omega_{\mathrm{DM}} h^2 = 0.111 \pm 0.006.
\end{equation}
We calculate the annihilation cross section for the $A_H$ in this
littlest Higgs model with T-parity, and evaluate the resulting
relic abundance as a function of the breaking scale $f$, and
$m_H$.  Annihilation to leptons and quarks occurs through either
s-channel Higgs exchange, or T-odd fermion doublet exchange diagrams,
both of which are suppressed due to smallness of the couplings.  The exception to this is the
annihilation to top quarks, which has a T-channel $t_-$ singlet exchange
diagram that is not suppressed.  Additionally, there are effective
Higgs-gluon-gluon and Higgs-photon-photon vertices coming from
loop diagams, but the annihilation cross section to these final
states is negligible.  The dominant contribution to the
annihilation coefficient for most considered values of $m_H$ and
$f$ arises from $W^\pm$ production through Higgs exchange.  The
production of $Z$-bosons is also similarly sizable, although
approximately a factor of $4$ smaller. The annihilation to Higgs
pairs is also quite large when allowed by phase space, and
dominates over annihilation to $Z$ bosons, though the $W^\pm$
channel still gives the largest contribution. The dominant
diagrams from the primary channels are shown in Figure
~\ref{annchan}.

\begin{figure}[t]
\centerline{\includegraphics[width=0.7\hsize]{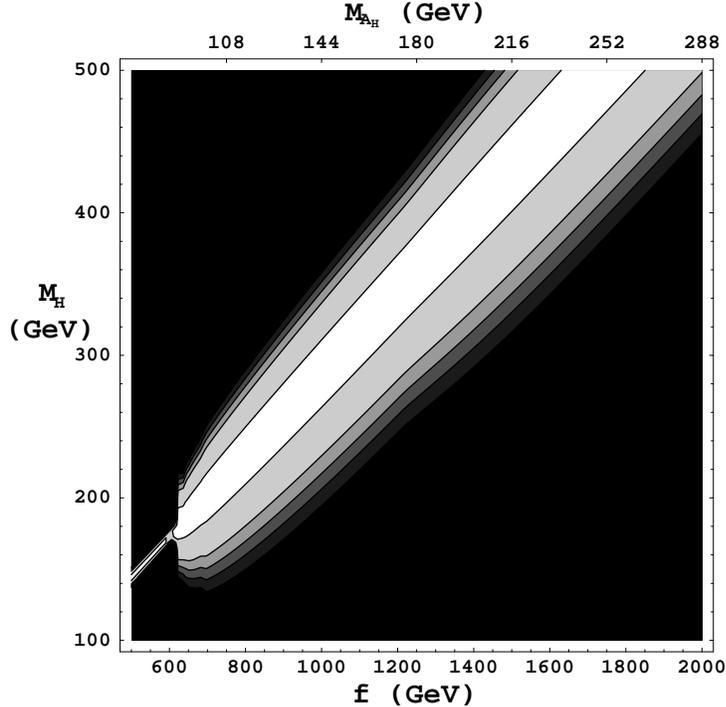}}
\caption{This plot depicts the variation of the relic density with
  respect to the Higgs mass and the symmetry breaking scale, $f$.
  In order from lightest to darkest regions, the $A_H$ makes up
  ($0-10\%$, $10-50\%$, $50-70\%$, $70-100\%$,
  $100\%$, $> 100\%$) of
  the observed relic abundance of dark matter.}
\label{limit-fig}
\end{figure}

There are regions of parameter space in which the $A_H$ is nearly
equal to half the mass of an s-channel exchanged particle.  In
this scenario, there are s-channel poles in the annihilation cross
section, and the diagrams which include such exchanges dominate
the cross section.  In this case, the annihilation rate is given
by
\begin{equation}
\sigma_A v \approx \frac{\gamma^2 s}{(m^2-s)^2 + m^2 \Gamma^2}
\end{equation}
where $\Gamma$ is the decay width of the exchanged particle, $s$
is the center of mass energy squared, and  $\gamma^2$ is a
prefactor that is dependent on the couplings of the $A_H$ to the
exchanged particle. This is quite important in this model, due to
the lightness of the $A_H$ in comparison with the breaking scale,
$f$. Many of the annihilation diagrams involve s-channel Higgs
exchange, so when the Higgs has twice the mass of the heavy
photon, the cross section will become quite large.

The resulting relic density is plotted in Figure
\ref{limit-fig}.\footnote{We are grateful to Maxim Perelstein and
  Andreas Birkedal for pointing out a factor of 4 error in our
  original calculation of the relic density.}
It is conceivable that there is another relic in addition to the
little Higgs dark matter, so we do not consider as ruled out
regions where the $A_H$ does not account for all of the dark
matter. In the black regions, there is too much dark matter left
over.  This is generically a worse scenario, since it would
overclose the universe, and we consider these regions to be ruled
out if the heavy photon is stable.  Interestingly, we find that
small values of the Higgs mass are disfavored if the $A_H$ is
indeed the WIMP.  Looking at Figure \ref{limit-fig}, one sees the
importance of the s-channel Higgs exchange along the line $m_h = 2
m_{A_H}$.  Along this contour, the pole in the annihilation
amplitude dominates the behavior of the annihilation cross
section.

There are regions of parameter space where
standard model particles are slightly heavier than the $A_H$, but there are still
$A_H$ particles on the high velocity end of the Boltzmann
distribution, and which are thus energetic enough to be able to
pair produce these slightly heavier particles, thus slightly
increasing the thermally averaged cross section just below thresholds.  In general, taking
these corrections
into account will smooth out the thermally averaged cross section
as the mass of the $A_H$ approaches such annihilation thresholds.
It does strongly affect the overall
fit, thus we neglect the threshold corrections in this paper.

The steep gradient of the relic density at $M_{A_H} \approx 80$
GeV is due to the threshold for annihilating to standard model $W$
bosons. Below $80$ GeV, the only available channels are to light
fermions. These channels have very small associated amplitudes, as
they require either the s-channel Higgs exchange which is suppressed by Yukawa couplings, or $T$-channel
T-odd fermion doublet exchange.  The $T$-channel fermion exchange diagrams are suppressed since the
relevant couplings $A_H \Psi_\mathrm{SM} \Psi_-$ are given by $g'/10
\approx .03$.

In the model that we have outlined, the strongest search
constraints would come from nuclear recoil experiments and high
energy solar and terrestrial neutrino searches.  Other
astrophysical searches, such as anomalous cosmic ray searches,
would not likely be fruitful. This is because the dominant channels for such
events require t-channel exchange of the heavy fermions, which, as
mentioned above, involve small couplings $g'/10$,
suppressing the relevant cross sections.  In nuclear recoil
experiments, however, because of the high density of gluons in
large nuclei, there is an enhancement of the nuclear scattering
cross section from the effective higgs-glue-glue vertex.  High
energy neutrino searches rely on gravitational capture of the
WIMP.  From the same argument, the nuclear scattering cross
section is sizeable, and serves to slow down WIMPs which encounter
the sun or earth.  The WIMPs are then captured gravitationally by
the sun or earth.  Subsequent annihilation of the $A_H$ will
potentially produce high energy neutrinos that reach detectors.
We leave a full analysis of the dark matter search constraints for
future research.

\section{LHC Collider Phenomenology}
\label{collider} \setcounter{equation}{0} \setcounter{footnote}{0}

The introduction of T-parity to the littlest Higgs adds new
interesting features and avoids problems of other little Higgs
models, but one must also check to see if there is any chance of
detecting its consequences in future collider experiments.  With
T-parity one loses the ability to singly produce the new heavy
vector bosons and scalars which is a major feature of how the
original littlest Higgs could be detected~\cite{maxim1,han1}. This
is not entirely disheartening since the signal of the littlest
Higgs will now become similar to that of the MSSM with a missing
energy signal, which has been analyzed in great detail. In fact
with the addition of T-parity type models it is even more
difficult to determine what type of new physics is discovered at
the LHC, since there are now more ways to fake a SUSY signal than
there were before~\cite{UED}.

The littlest Higgs with T-parity has more varied collider
signatures than just a missing energy signal which could fake
SUSY. Not all of the new particles in the littlest Higgs with
T-parity are T-odd.  As in the original littlest Higgs model there
is a T-even partner of the top quark which can be singly produced.
The key difference in the littlest Higgs with T-parity compared to
previous studies of the T-even partner of the top quark is the
existence of the T-odd partner of the top quark, $t'_{-}$.  The $t'_{-}$ in fact will always be
lighter than the $t'_{+}$~(see (\ref{tevenmass}) and
(\ref{toddmass})). This will open up new decay channels for the
$t'_{+}$ that did not exist in previous studies.  These new decay
channels can contribute a sizeable branching fraction to the
$t'_{+}$ and thus new studies of the T-even sector of this model
as well as the T-odd sector are required.

In this section we start the analysis of the collider
phenomenology of the T-parity littlest Higgs.  We choose sectors which capture the
phenomenology of the scalar, gauge boson, and heavy fermion sectors
that are common to all implementations of the T-parity littlest
Higgs.  We must emphasize here that a more general study including the
additional T-odd fermions besides the $t'_-$ should be undertaken,
especially considering that the production cross sections involve
strong rather than weak couplings.  In
Section~\ref{toddprod} we will investigate the production
mechanism for the T-odd particles in this model and the branching
fractions for their decays.  In Section~\ref{teventop} we will
revisit the production and decay channels of the heavy T-even top
that is now changed due to the presence of $t'_{-}$.  We will then
discuss in Section~\ref{backgrounds} the best search mechanisms as
well as the associated backgrounds.

\subsection{T-odd phenomenology}\label{toddprod}

\begin{figure}[ht]
\centerline{\includegraphics[width=0.6\hsize]{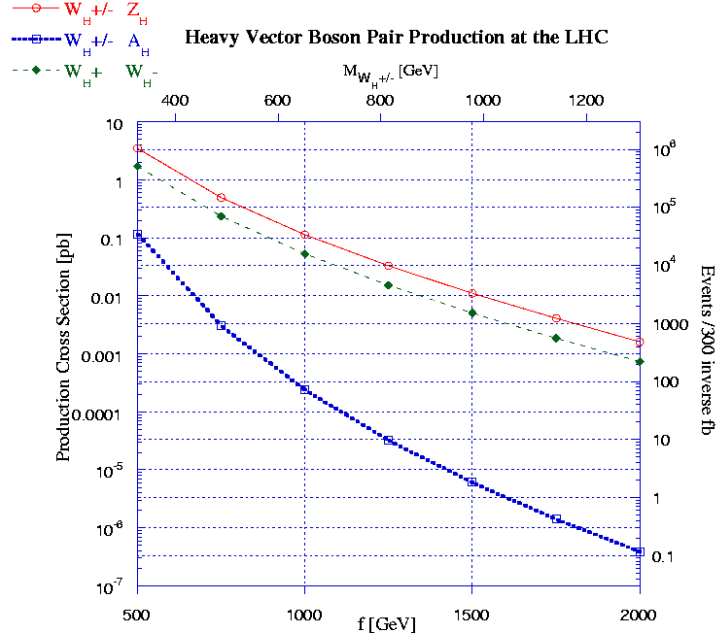}}
\caption{The cross section for the production of a pair of T-odd
heavy vector bosons at the LHC is plotted as a function of the
symmetry breaking scale $f$.  The number of events for
$300\;\mathrm{fb}^{-1}$ is plotted on the second y-axis.
$M_{W_H^{\pm}}$ is plotted on the second x-axis. $M_{Z_H}$ is
degenerate in mass with $M_{W_H^{\pm}}$, and $M_{A_H}\sim .16 f$.
} \label{vectorproduction}
\end{figure}

We will begin our study with the cross sections for pair producing
T-odd particles.  Implementing the model in COMPHEP~\cite{comphep}
we compute production cross sections for the LHC.  In the heavy
gauge boson sector of the model there are no free parameters other
than the scale $f$ so the production cross sections at leading
order are unambiguous.  The cross section for pair producing heavy
gauge bosons is plotted in Figure~\ref{vectorproduction}.
For proton-proton scattering at the LHC the dominant production
channels are $W_H^{\pm} Z_H$ or $W_H^{+} W_H^{-}$ pairs, while
other channels such as $W_H^{\pm} A_H$ have lower total cross
sections. For $W_H^{\pm} Z_H$ and $W_H^{\pm} A_H$ pairs, the
production is from the exchange of a $W^{\pm}$; however the
$W_H^{\pm} A_H$ pairs interaction with the SM $W^{\pm}$ is $v/f$
suppressed.  The $W_H^{+} W_H^{-}$ pairs are produced through the
exchange of a photon or $Z$.  The decay channels of the heavy
gauge bosons are simple since it turns out that they always decay
directly to $A_H$, the lightest T-odd particle.  The $Z_H$ decays
exclusively to $A_H\, h$ and the $W_H^{\pm}$ decays entirely to
$A_H\, W^{\pm}$.

The heavy triplet $\Phi$ is also T-odd and must be pair produced.
The various components of the triplet
$\phi^{++}$, $\phi^+$, $\phi^0$, $\phi^P$ and their antiparticles all
have the same mass at tree level.  The mass of the $\Phi$ is
related to the mass of the Higgs through the
relation~(\ref{tripletmass}).
\begin{figure}[ht]
\centerline{\includegraphics[width=0.6\hsize]{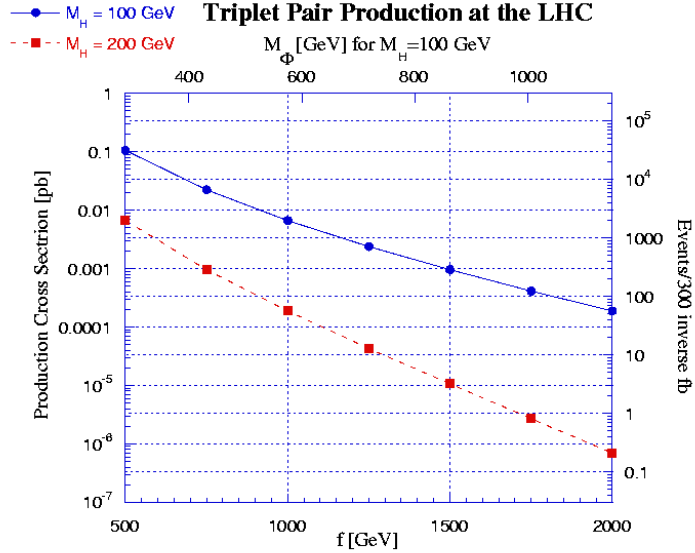}}
\caption{The cross section for the production of a pair of T-odd
triplets at the LHC is plotted as a function of the symmetry
breaking scale $f$.  The cross section is plotted for $m_H=100,200
\;\mathrm{GeV}$ since the triplet mass, $M_\Phi$, is determined by
$f$ and $m_H$. The number of events for $300\;\mathrm{fb}^{-1}$ is
plotted on the second y-axis.  $M_\Phi$ for a Higgs mass of
$100$~GeV is plotted on the second x-axis, for a Higgs mass of
$200$~GeV simply scale the second x-axis by a factor of
$2$.}\label{tripprodfig}
\end{figure}
In principle one should analyze all the different production
channels for the components of $\Phi$ individually however to get
an overall idea of the magnitude for pair producing $\Phi$'s we
sum the contribution for all channels and plot the cross section
for the LHC in Figure~\ref{tripprodfig}.  In
Figure~\ref{tripprodfig} since the mass of the $\Phi$ is
determined by $m_H$ and $f$ we plot the production cross section
as a function of $f$ for two different values of $m_H$.  The
dominant channels for production of the triplet, in the naive
scenario where all components of the triplet have the same mass,
are for the charged components of $\Phi$ from $W^{\pm}$ exchange.
The decay of the components of the triplet are as simple as for
the heavy gauge bosons since they each have only one decay
channel. The charged components of the triplet decay in the
following ways, $\phi^{++}\rightarrow W^+ W_H^+$ and $\phi^+
\rightarrow A W_H^+$, with corresponding decay channels for the
anti-particles. The pseudoscalar $\phi^P$ and scalar $\phi^0$
decay through the processes $\phi^P\rightarrow H A_H$ and
$\phi^0\rightarrow Z A_H$.

\begin{figure}[h]
\centerline{\includegraphics[width=0.6\hsize]{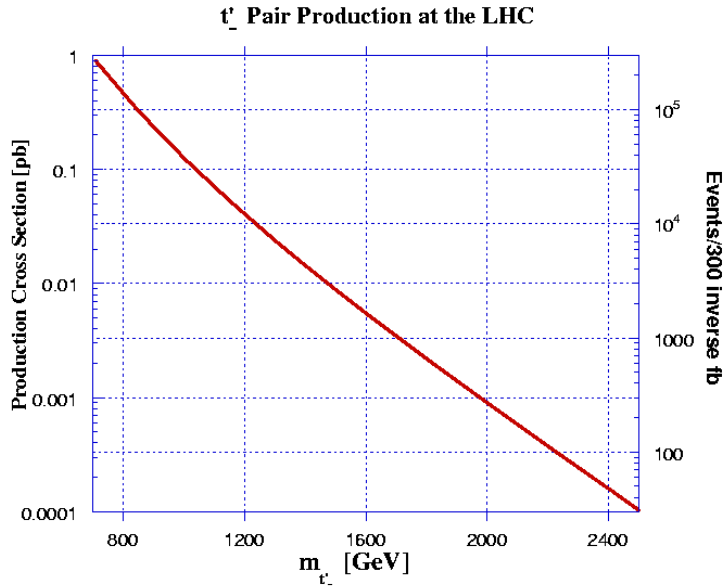}}
\caption{The cross section for the production of a pair of T-odd
heavy quarks $t'_{-}$ at the LHC is plotted as a function of
$m_{t'_{-}}$.  The number of events for $300\;\mathrm{fb}^{-1}$ is
plotted on the second y-axis.}\label{tminusfig}
\end{figure}

The production process for the $t'_{-}$ turns out to be very
similar to pair producing $t'_{+}$ as in~\cite{han1}.  The reason
why the production is the same as the $t'_{+}$ in this particular
channel is that the production cross section is dominated by gluon
exchange and is independent of all parameters except for
$m_{t'_{-}}$.  One should keep in mind that even though cross
sections of $t'_{-}$ and $t'_{+}$ have the same mass dependence,
$t'_{-}$ and $t'_{+}$ have in general different masses so they are
produced at different rates.  In Figure~\ref{tminusfig} we plot
the cross section for producing the $t'_{-}$ at the LHC as a
function of its mass.
The decay pattern of the $t'_{-}$ is simple, the $t'_{-}$ decays
entirely to $A_H\,t$.  The phenomenology of the $t'_{-}$ merits
further study since it will always be lighter than the $t'_{+}$ as
well as having a large cross section for production.  There may be
other interesting channels for decay if one examines a different
spectrum for the model. Since the cross section for producing the
$t'_{-}$ is so large it may be one of the more interesting ways to
search for T-parity, however the backgrounds must be considered
which will be done in Section~\ref{backgrounds}.

\subsection{T-even Top Quark}\label{teventop}

The main difference between R-parity SUSY models, the UED models
with KK-parity that could fake a R-parity SUSY signal~\cite{UED},
and T-parity, is the existence of the $t'_{+}$. The $t'_{+}$ is a
generic feature of any little Higgs model and also any known model
of T-parity. Comparing the interactions of $t'_{+}$ in the
T-parity littlest Higgs to the original littlest Higgs
model~\cite{littlest}, one finds that the same interactions
dominate the production cross section.  The largest cross section
for producing the $t'_{+}$ at the LHC is for singly producing a
$t'_{+}$ and a jet, from T-channel $W$ exchange.  For low $t'_{+}$
mass a comparable cross section is for pair producing $t'_{+}$
however it drops rapidly with increasing mass.  For a plot of the
production cross section for the $t'_{+}$ we refer the reader
to~\cite{han1,maxim2} since the results are the same for this
model.

\begin{figure}[ht]
\centerline{\includegraphics[width=0.6\hsize]{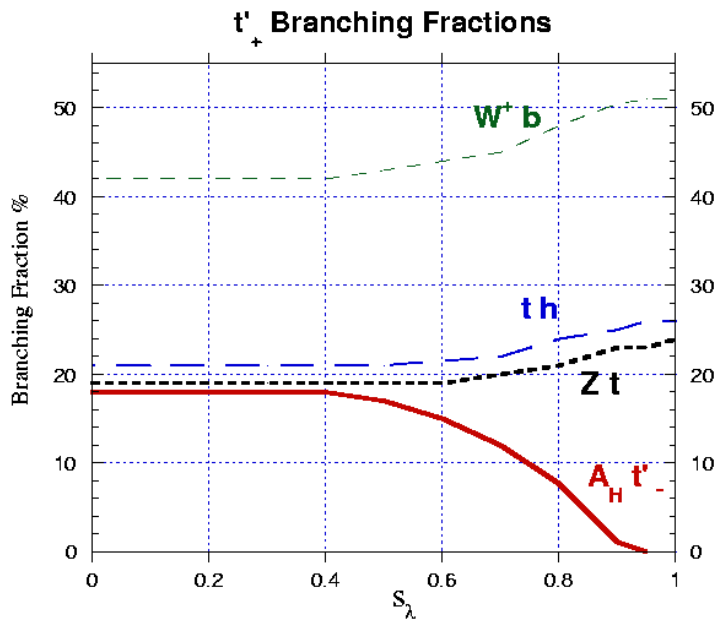}}
\caption{We plot the branching fractions for $t'_{+}$ decay as a
function of $s_{\lambda}$, which parameterizes the ratio of masses
of the $t'_{+}$ and $t'_{-}$.  This plot was generated for
$f=1$~TeV}\label{tbranch}
\end{figure}

Even though the production cross sections for the $t'_{+}$ are the
same in T-parity models as for non T-parity models at the LHC,
there is an important difference in the phenomenology of the
$t'_{+}$.  The existence of the $t'_{-}$ which is always required
to be less massive then the $t'_{+}$ from (\ref{toddmass}) and
(\ref{tevenmass}) opens a new decay channel for the $t'_{+}$. In
Figure~\ref{tbranch} we plot the branching fractions of the
$t'_{+}$ as a function of
$s_{\lambda}=\frac{m_{t'_{-}}}{m_{t'_{+}}}$. The branching
fraction is essentially independent of $f$.  As one can see from
Figure~\ref{tbranch}, for most of parameter space the $t'_{+}$ has
a sizeable invisible width from decay to $t'_{-} A_H$.  In reality
though to solve the little hierarchy problem one is only
interested in the region around $s_\lambda=\frac{1}{\sqrt{2}}$
where $m_{t'_+}=\sqrt{2}f$.  For either direction in $s_\lambda$,
$m_{t'_+}$ increases which causes a fine tuning of the Higgs mass
if $m_{t'_+}$ is larger than $\sim 2$~TeV.  The existence of this
new sizeable invisible width of the $t'_{+}$ does not let one
apply the analysis of \cite{maxim2} for the $t'_{+}$ in T-parity
models.  In \cite{maxim2} it was hoped that one could test the
little Higgs mechanism for the $t'_{+}$ by measuring the couplings
of the $t'_{+}$ and $f$ independently, since they must satisfy a
particular relationship to cancel the one-loop quadratically
divergent contributions to the Higgs mass from the top quark. In
T-parity type models one first has a difficulty with measuring $f$
because one cannot obtain $f$ from the gauge boson sector as in
\cite{maxim1}. In addition, the new sizeable partial width of the
$t'_{+}$, which is hard to determine, makes measuring the
couplings of the heavy partner of the top quark at the LHC virtually
impossible.

\subsection{Backgrounds and Best Signals}\label{backgrounds}

In the previous parts of this section we have studied the
tree-level production cross sections for the LHC and found a large
number of events for certain processes.  However when dealing with
a hadron collider the background tends to be ubiquitous so the
question remains, for the interesting processes can one
distinguish signal from background?  To begin analyzing this
question we compute the tree level production cross sections for
irreducible backgrounds at the level of SM gauge and Higgs bosons.
This analysis is not at the level of a physical background in a
detector, however a signal must pass this simple test before
taking into account the full background.  For those processes
where we find the signal is much higher than the simple
backgrounds we analyze, we comment on the full backgrounds.  A
full analysis of the backgrounds is required but we leave this for
future work.

We will start by looking at the backgrounds for the pair
production of heavy gauge bosons.  From Section~\ref{toddprod} the
strongest production mechanisms for the heavy gauge bosons are in
the channels $W_H^{\pm} Z_H$ and $W_H^{+} W_H^{-}$.  Since the
heavy gauge bosons have only one decay channel each, the final
state (after the heavy particles have decayed) will be
\begin{eqnarray}\label{vecdecay}
p\, p &\rightarrow& W_H^{\pm}\, Z_H \rightarrow W^\pm \,h\, A_H\, A_H \nonumber \\
p\, p &\rightarrow& W_H^{+}\, W_H^{-} \rightarrow W^+\, W^-\,
A_H\, A_H.
\end{eqnarray}
The signature in the detector will then be missing energy from the
$A_H$ along with the decay products of the SM particles.  For the
simple irreducible background we are interested in, we look for SM
processes which have the same SM particles (at the level of gauge
and Higgs bosons) as (\ref{vecdecay}) along with missing energy in
the final state. The dominant process of this type in the SM are
to include a $Z$ in the final state in place of the $A_H$'s, which
then decays to neutrinos. Thus our estimate will be to compute the
production cross section at the LHC for the final states of
(\ref{vecdecay}) replacing the $A_H$'s with a single $Z$ and then
multiplying by the branching fraction for the invisible decay of
the $Z$.

For $W_H^{+} W_H^{-}$ production one has to compete with a
background of triple gauge boson production $W^+ W^- Z$ which has
a production cross section, taking into account the $Z$ branching
fraction, of $\sim 10^{-2}$ pb.  Comparing this background rate to
the signal found in Fig.~\ref{vectorproduction}, we find that the
signal is larger than the background for small values of $f$. To
determine whether the signal is actually observable, since the
signal passes this test for a certain range of $f$ one must also
consider the other processes which have the same final state in
the detector (not just at the level of gauge and Higgs bosons). In
the detector the possibilities for this channel are
\begin{eqnarray}
p\, p \rightarrow W_H^{+}\, W_H^{-} &\rightarrow&
4\;\mathrm{jets}\,+\,\mathrm{MET},\\
&\rightarrow& 2\;\mathrm{jets}\,+\, l\, +\,\mathrm{MET},\label{2jet}\\
&\rightarrow& 2\, l\, +\,\mathrm{MET}\label{2l}.
\end{eqnarray}
Ignoring the $4$ jets $+$ MET channel for its inherent
difficulties, unfortunately this type of signal ((\ref{2jet}) and
(\ref{2l})) in a detector has a background from
\begin{equation}
p\, p \rightarrow W^{+}\, W^{-}
\end{equation}
which has a production cross section of $\mathcal{O}(\mathrm{pb})$
at the LHC.  This background is significantly larger than the
signal and makes discovery in the $W_H^{+} W_H^{-}$ channel very
difficult.  Since the MET from the signal will be very energetic
with the application of cuts perhaps some signal can be
distinguished from background but it would be very difficult.

For the $W_H^{\pm} Z_H$ production channel our simple irreducible
background estimate comes from $W^\pm \, h\, Z$ production.  The
total production cross section for this background, taking into
account the branching fraction of $Z\rightarrow
\mathrm{invisible}$, is $\sim 10^{-3}$ pb. When comparing this
background rate to the signal using Fig.~\ref{vectorproduction} we
find that for all values of $f$ the signal is larger than the
background and for $f\sim 1\;\mathrm{TeV}$ the signal is almost
two orders of magnitude larger.  This is relatively easy to
understand since producing a Higgs with two gauge bosons is
difficult in the SM.  However, we must estimate a more realistic
background for this process as we did for the $W_H^{+} W_H^{-}$
channel, before deciding whether this is a promising channel. For
low Higgs mass the Higgs will decay predominantly into $b$ jets,
so to avoid looking for a $4$ jet plus MET signal, we consider the
$W$ decaying into a lepton plus MET. Therefore the backgrounds can
come from
\begin{eqnarray}
p\, p & \rightarrow & h\, W^{\pm} \rightarrow 2\;b\;\mathrm{jets}+l+\,\mathrm{MET},\\
& \rightarrow & Z\, W^{\pm} \rightarrow 2\;b\;\mathrm{jets}+l+\,\mathrm{MET},\\
\end{eqnarray}
which will have a rate comparable to the signal we are interested
in, making discovery in this channel also unlikely.  However further study
of this channel would be interesting, since it has the highest rate of production of
the heavy gauge bosons.

The production of the triplet $\Phi$ is even less promising than
the heavy gauge bosons since the production cross section is
lower, Fig.~\ref{tripprodfig}, and the backgrounds are more
complicated. In general the signal for pair producing components
of $\Phi$ will be the SM decay products of multiple gauge bosons
along with missing energy. The most promising component of the
triplet is the pseudoscalar $\phi^P$ which decays to $A_H\,h$,
since the Higgs is a more unique signal than SM gauge bosons.  The
two ways to produce a $\phi^P$ come from
\begin{eqnarray}
p\, p &\rightarrow& \phi^{\pm}\, \phi^P \rightarrow W^\pm \,A\, h\, A_H\, A_H \nonumber \\
p\, p &\rightarrow& \phi^0\, \phi^P \rightarrow h\, Z\, A_H\, A_H.
\end{eqnarray}
The cross section for the production channel $\phi^0\, \phi^P$ is
of the order $10^{-4}$ pb, for a Higgs mass of $115$ GeV and
$f=1$~TeV, while the simple irreducible background, which was
estimated as before by replacing the $A_H$ with a $Z$ decaying
invisibly, is of the same order.  Since the signal for $\phi^0\,
\phi^P$ is not larger than even the naive background we don't
consider the full background for this channel.  A slightly more
promising channel is the $\phi^{\pm}\, \phi^P$ which also has a
cross section for production of the order $10^{-4}$ pb for a Higgs
mass of $115$ GeV and $f=1$~TeV. The simple irreducible background
in this case would be a final state $W^\pm \,A\, h\, Z$ with the
$Z$ decaying invisibly. This background at tree level is of the
order $10^{-6}$ pb which seems promising.  However with so few
events from the signal, any enhancement of the background beyond
this naive estimation could prove troublesome.  The signal in the
detector for this process, assuming the Higgs decays to $b$ jets
and the $W$ to leptons, is
\begin{equation}
p\, p \rightarrow 2\;b\;\mathrm{jets}+l+A+\,\mathrm{MET}.
\end{equation}
The physical background for this signal could come from $W,A,h$
production.  The production rate for this background is comparable
to the signal so it does not appear promising.  However, the
photon would be very energetic coming from the decay of
$\phi^{\pm}$ so perhaps a cut could be made to reduce the
background.

The detection of the $t'_{+}$ in the littlest Higgs model without
T-parity has been extensively investigated for the LHC in a study
for the ATLAS detector~\cite{atlas}. The only difference for the
$t'_{+}$ when including T-parity is the sizeable new contribution
to its width from the $t'_{-}$ as discussed in
Section~\ref{teventop}.  The $t'_{-}$ is one of the most
interesting new parts of the T-parity littlest Higgs since it
doesn't exist in non T-parity little Higgs models, and it is
required to be lighter than the $t'_{+}$, which in turn is
required to be approximately at a TeV to avoid fine tuning. Even
though the $t'_{-}$ is perhaps the most interesting new particle
from the model building perspective, it will be difficult
to discern at the LHC.  The $t'_{-}$ decay pattern gives a
background at the LHC of $t\, \bar{t}\,+\,\mathrm{MET}$.  The
production rate for $t\, \bar{t}\,+\,\mathrm{MET}$ at the LHC is
large, making it difficult to find this signal above
background.

\section{Conclusions}
\label{conclusions-sec}
\setcounter{equation}{0}
\setcounter{footnote}{0}

In this paper we have given a review of the implementation of
T-parity in the littlest Higgs model and begun the study of its
phenomenology. There are several implementations of T-parity in
the littlest Higgs as discussed in Section~\ref{alternatives}. We
studied the phenomenology of a particular implementation that has
the same basic features of any implementation of littlest Higgs
with T-parity. In studying the dark matter candidate of the
littlest Higgs with T-parity (if T-parity is conserved), which is
the heavy photon $A_H$, we find that it can account for the
observed relic density of dark matter in the universe.  Assuming
that the $A_H$ is stable, we find bounds on the symmetry
breaking scale $f$ as a function of $m_H$, and that lower values
of $m_H$ are disfavored. We have begun the study of the collider
phenomenology of the littlest Higgs with T-parity. We find that
the generic signal for this model will be a missing energy signal
similar to that of SUSY. The littlest Higgs model with T-parity
offers a distinct alternative to SUSY for stabilizing the
electroweak scale, and could be thought of as a ``bosonic"
supersymmetry since it could fake a SUSY signal at the LHC.  We
find that since one must pair produce T-odd particles the signal
will be very difficult to distinguish from the background, which
is very different than the littlest Higgs model without
T-parity~\cite{maxim1,han1}. The T-parity littlest Higgs also has
a heavy T-even top quark partner $t'_{+}$ which distinguishes
itself from a supersymmetric model or one designed to fake a
supersymmetric signal.  The production mechanisms of the $t'_{+}$
turn out to be identical to the littlest Higgs without T-parity,
however we find that in introducing T-parity one is required to
introduce a T-odd $t'_{-}$ with a mass required to be lighter than
that of the $t'_{+}$. The $t'_{-}$ changes the phenomenology of
the $t'_{+}$, and thus existing studies of the $t'_{+}$ are not
directly applicable for determining its properties.

There are many avenues for future research on the littlest Higgs
with T-parity and we list some of the most important in our
opinion. We have found bounds on the scale $f$ from the
analysis of the dark matter candidate (under the assumption that
T-parity is conserved), however, a full analysis of EW precision
constraints including loop contributions needs to be done to
 find complementary constraints.  The discovery channels that we have discussed need to be further
investigated beyond our simple treatment of the background to see
if this model can really be distinguished or not from the
ubiquitous background of the LHC.   One might also investigate the
T-odd fermion spectrum, which is quite interesting from the standpoint
of phenomenology.  We analyzed only the gauge, scalar, and top sectors,
however in principle all the new fermions could be within reach of
the LHC making the parameter space much larger and give other
opportunities for discovery.  A study of how to distinguish the
littlest Higgs with T-parity from the MSSM (or other models
designed to fake SUSY) at the LHC is also of paramount interest.
The $t'_{+}$ would be an obvious distinguishing characteristic of
T-parity however it could be out of the discovery reach of the
LHC.  Although the existence of T-parity greatly complicates
deciphering the data of the LHC, it provides an interesting
solution to the little hierarchy problem that deserves further
study.

\section*{Acknowledgments}

We would like to thank Maxim Perelstein for useful discussions as
well as reading the manuscript and making helpful comments.  We
would like to thank Ian Low for clarification of some aspects of
the T-parity models. We would also like to thank Csaba Cs\'{a}ki
and Konstantin Matchev for reading the manuscript and making
helpful comments. J.H. and P.M would like to thank the University
of Colorado at Boulder TASI program, and P.M would like to thank
the T-8 group at Los Alamos National Lab for hospitality and
support while this work was in progress. The research of J.H. and
P.M. is supported in part by the NSF under grant PHY-0355005.

\appendix

\section{Four Fermion Operators}\label{app:fourfermi}

\begin{figure}[ht]
\centerline{\includegraphics[width=0.8\hsize]{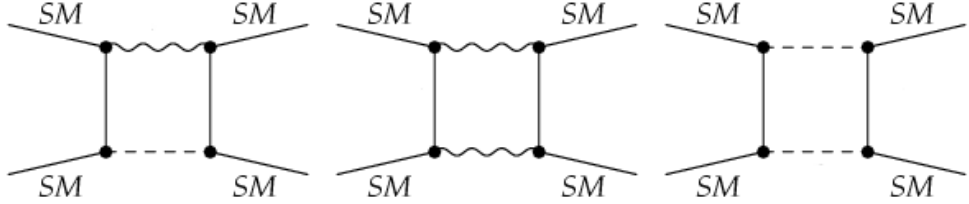}}
\caption{These diagrams, where the particles running in the loop
are
  all odd under T-parity, lead to potentially dangerous
  four-fermion operators.  In these diagrams, the scalar propagators
  represent pion exchange, and the gauge boson propagators refer to
  exchange of any of the $A_H$, $Z_H$, or $W_H^\pm$.  The fermion
  lines in the loops represent exchange of the heavy T-odd
  fermions.}\label{4fdiag}
\end{figure}

The four fermion operators that ruled out the models of
\cite{tpar2} are absent in this model, since the standard model
fermions transform linearly.  However, there are still finite
diagrams which are potentially dangerous, and must be checked.  In
the model that we discuss, the types of diagrams are shown in
Figure \ref{4fdiag}.  By NDA, these diagrams give effective
four-fermion contact terms which are approximately given by:
\begin{equation}
\frac{g^4 \kappa^2}{16 \pi^2 \lambda^2 x^2 f^2}, \ \ \frac{g^4}{16
\pi^2} \left(\frac{1}{\lambda f}\right)^2\ \ \mathrm{and} \ \
\frac{\kappa^4}{16 \pi^2} \left(\frac{1}{\lambda f}\right)^2,
\end{equation}
respectively.  The parameter $x$ refers to the ratio of the pion mass
to the breaking scale, $f$.  We have taken $x$ to be larger than $g$.  In order to avoid
bounds on such contact terms, these must be less than $1/(5-10
\mathrm{TeV})^2$.

One might expect that these can be easily suppressed by taking
$\lambda$ to be large, however a more careful analysis is
required. The interactions involving the coupling $\kappa$
originate from the heavy fermion Yukawas
\begin{equation}\label{heavyyuk2}
\kappa f (\bar{\Psi}_2 \xi \Psi'+\bar{\Psi}_1 \Sigma_0 \Omega
\xi^\dagger \Omega \Psi').
\end{equation}
where the fermion multiplets are given by
\begin{equation}
\begin{array}{ccc}
\Psi_1=\left(\begin{array}{c} \psi_1 \\ 0 \\ 0 \end{array}\right)
& \Psi_2=\left(\begin{array}{c} 0 \\ 0 \\ \psi_2
\end{array}\right) &
\Psi'=\left(\begin{array}{c} \tilde{\psi}' \\ \chi' \\ \psi'
\end{array}\right)
\end{array}.
\end{equation}
It is from this term that the diagrams with Goldstone boson
exchange in the loop emerge.  However, this term is also
generating a mass term $\sqrt{2} \kappa$ for the Dirac pair $(
\psi_2 + \psi_1 , \psi' )$.  This could potentially give a fixed
relation between $\kappa$ and $\lambda$ that would force the
parameter $\kappa$ to be relatively small, prohibiting a
decoupling limit.  For example, if the mass of the heavy fermion
is $\sqrt{2} \kappa$, then the contact operator would be
\begin{equation}
\frac{\kappa^4}{32 \pi^2} \left(\frac{1}{\kappa f}\right)^2 =
\frac{\kappa^2}{32 \pi^2 f^2}.
\end{equation}
In this case, we would require $\kappa < 0.5 - 1.5$ for $f=1$ TeV.

Most of the interactions in (\ref{heavyyuk2}), however, do not lead to
amplitudes where we have such a relation, and many of the fermion
masses can be made quite heavy.  In particular, it is only the right
handed doublet, $\psi'$ which must be kept close to the breaking scale
$f$.  The four-fermion operator that involves $\psi'$ exchanges the
Goldstones which are eaten by the heavy vector fields $W_H$ and
$Z_H$, and these operators can not be suppressed since it is $\psi'$
which gets its mass from the Yukawa interaction in
Eq. (A.2).\footnote{We thank Thomas Gregoire for pointing out that
  this is the relevent four fermion diagram that prevents decoupling
  the T-odd fermion spectrum completely.}  All of the vertices
appearing in Figure \ref{4fdiag} that involve the triplet, $\Phi$ couple
the standard model fermion to the doublet $\tilde{\psi}'$, rather
than $\psi'$.  The parameters $\lambda$ that appear in the four
fermion diagrams above are then independent of $\kappa$.  Simply
taking $\lambda > \kappa^2$ suppresses the dangerous amplitudes.
We note, though, that we do not take $\lambda$ to the strong
coupling limit.  First, this would be dangerous for potential
quartic divergences in the Higgs mass \cite{tpar2}. Second, we are
attempting to avoid taking any parameters in the theory into a
non-perturbative regime.

We have shown here that we can in fact raise the masses of the
many of the additional fermions above the minimal little Higgs spectrum.  In
example, for an $f$ of $1$ TeV, the masses of most T-odd fermions
are all at $5$ TeV, safe enough to neglect in collider and dark
matter phenomenology, yet light enough to avoid theoretical issues
involving large contributions to the Higgs mass through two loop
quartically divergent diagrams.  The couplings required to raise
the masses sufficiently are still within the perturbative regime,
so we have not resorted to a strong coupling limit.  With the
resulting TeV scale spectrum, we compute the phenomenology of a
littlest Higgs with T-parity which is similar to the low energy
limit of the other models which incorporate this discrete
symmetry.  We leave an analysis of the lighter T-odd partners of the
standard model doublets for future research.

\section{Feynman Rules for the Littlest Higgs with
T-parity}\label{app:feyn}

In this appendix we will provide a list of Feynman rules specific
to the T-parity littlest Higgs as a reference to facilitate
further study of the model.  As discussed in
Section~\ref{sec:model} implementing T-parity in the littlest
Higgs for the gauge and scalar sectors does not require a drastic
change to the structure of the original littlest Higgs. The effect
in the gauge sector is to set the gauge couplings, $g_j$, in the
original littlest Higgs to be $g_1=g_2=\sqrt{2} g$, and likewise
for the $g'$ couplings. In the scalar sector the effect is
essentially to forbid a VEV for the triplet $\Phi$ by forbidding
the T-odd coupling $H\Phi H$.  With this in mind a great many of
the Feynman rules can be obtained from~\cite{han1} which listed
all the Feynman rules for the littlest Higgs.  We will give a
guide to how to obtain the T-parity Feynman rules from the Feynman
rules in~\cite{han1}. We then will list the additional Feynman
rules specific to the T-parity littlest Higgs.

\begin{table}[h]
\center{\begin{tabular}{|c|c||c|c|}
\hline Particles & Vertices & Particles & Vertices \\

\feyn {\whm{\mu} \wlp{\nu} \phs} {\frac{g^2}{3 \sqrt{2}}
\frac{v^2}{f} \met} {Z^\mu \zh{\nu} \pho} {-i \frac{g^2}{2\sqrt{2}
\cw} \frac{v^2}{f} \met}

\feyn {\ah{\mu} Z^\nu \pho} {i \frac{g'^2}{2 \sqrt{2} \sw}
\frac{v^2}{f} \met} {\whp{\mu} Z^\nu \phm} {i \frac{g^2}{12 \cw} \left( 1+
2\sw^2\right) \frac{v^2}{f} \met}

\feyn {\whp{\mu} A^\nu \phm} {-i \frac{ e g}{6} \frac{v^2}{f}
\met} {\wlp{\mu} \ah{\nu} \phm} {-i \frac{g g'}{4} \frac{v^2}{f}
\met}

\feyn {\zh{\mu} \wlp{\nu} \phm} {i \frac{5 g^2}{12} \frac{v^2}{f}
\met} {\wlm{\mu} \whm{\nu} \phpp} {-i \frac{g^2}{2} \frac{v^2}{f}
\met}

\hline \feyn {\whm{\mu} \wlp{\nu} \phi^P h} {\frac{\sqrt{2}
g^2}{3} \EPS \met} {Z^\mu Z_H^\nu \phi^0 h} {-i
\frac{g^2}{\sqrt{2} \cw} \EPS \met}

\feyn {\ah{\mu} Z^\nu \pho h} {i \frac{g'^2}{\sqrt{2} \sw} \EPS
\met} {\whp{\mu} Z^\nu \phm h} {i \frac{g^2}{6 \cw} \left( 1+2
\sw^2 \right) \EPS \met}

\feyn {\whp{\mu} A^\nu \phm h} {-i \frac{e g}{3} \EPS \met}
{\wlp{\mu} \ah{\nu} \phm h} {-i \frac{g g'}{2} \EPS \met}

\feyn {\zh{\mu} \wlp{\nu} \phm h} {i \frac{5 g^2}{6} \EPS \met}
      {\wlp{\mu} \whp{\nu}
\phmm h} {-i g^2 \EPS \met}

\feyn {\zh{\mu} \zh{\nu} \pho \pho} {i \frac{g^2}{12}
\frac{v^2}{f^2} \met} {\ah{\mu} \ah{\nu} \pho \pho} {i
\frac{g'^2}{12} \frac{v^2}{f^2} \met}

\feyn {\zh{\mu} \ah{\nu} \pho \pho} {-i \frac{g g'}{12}
\frac{v^2}{f^2} \met} {\whp{\mu} \ah{\nu} \pho \phm} {i \frac{x_h
g^2}{15\sqrt{2}\cw^2} \left( 10-9 \sw^2\right)
  \frac{v^2}{f^2} \met}

\feyn {\whp{\mu} \whm{\nu} \php \phm} {i \frac{g^2}{24}
\frac{v^2}{f^2} \met} {\ah{\mu} \ah{\nu} \php \phm} {i
\frac{g'^2}{24} \frac{v^2}{f^2} \met}

\feyn {\zh{\mu} \ah{\nu} \php \phm} {i \frac{x_h g^2}{30 \cw^2}
\left( 25- 18 \sw^2\right) \frac{v^2}{f^2} \met} {\whp{\mu}
\ah{\nu} \php \phmm} {-i \frac{ x_h g^2}{15 \cw^2} \left( 5-3
\sw^2 \right) \frac{v^2}{f^2} \met}

\feyn {\zh{\mu} \zh{\nu} \phs \phs} {i \frac{g^2}{12}
\frac{v^2}{f^2} \met} {\ah{\mu} \ah{\nu} \phs \phs} {i
\frac{g'^2}{12} \frac{v^2}{f^2} \met}

\feyn {\zh{\mu} \ah{\nu} \phs \phs} {-i \frac{g g'}{12}
\frac{v^2}{f^2} \met} {\whp{\mu} \ah{\nu} \phs \phm} {- \frac{ x_h
g^2}{15 \sqrt{2} \cw^2} \left(10 -9 \sw^2\right) \frac{v^2}{f^2}
\met} \hline

\feyn {\whp{\mu} h \phm} {i \frac{ g}{6} \EPS \moms} {\ah{\mu} h
\phs} { \frac{g'}{3 \sqrt{2}} \EPS \moms}

\feyn {\zh{\mu} h \phs} {-\frac{g}{3 \sqrt{2}} \EPS \moms} {} {}

\hline
\end{tabular}}
\caption{Feynman rules for the gauge and scalar sector of the
T-parity littlest Higgs that can not be determined from the
appendix of~\cite{han1}.  The momenta are all defined as outgoing
and for the gauge-scalar-scalar vertices, the momenta refer to the
first and second scalar respectively.}\label{tparityzero}
\end{table}

If one uses Tables I-VIII in~\cite{han1} for the Feynman rules in
the scalar and gauge sectors, it is necessary to apply the
following rules for the parameters in~\cite{han1} to obtain the
correct vertices for the T-parity littlest Higgs model:
\begin{eqnarray}\label{taorules}
c&=&s=c'=s'=\frac{1}{\sqrt{2}}, \\
s_0&=&s_P=s_+=0,\\
c_0&=&c_P=c_+=1.
\end{eqnarray}
However when using (\ref{taorules}) with the tables referred to
in~\cite{han1} certain T-even interactions will naively be set to
$0$ since T-parity will cause certain interactions to start at a
higher order in a $v/f$ expansion than taken into account. These
interactions that can not be found in~\cite{han1} need to be
included for studying certain phenomenological processes.  As an
example certain components of the $\Phi$ would be unnaturally long
lived since most decay channels are missing if one does not
include the higher order interactions. In Table~\ref{tparityzero}
we list the Feynman rules for T-parity even interactions that are
zero in the order considered by~\cite{han1}.  The parameters
$s_\mathrm{W}$ and $c_\mathrm{W}$ refer to the sine and cosine of
the weak mixing angle $\theta_\mathrm{W}$, while
\begin{equation}
x_h=\frac{5}{4} \frac{g g'}{5 g^2-g'^2}.
\end{equation}

The fermion sector of the littlest Higgs with T-parity is
radically different than the original littlest Higgs model,
therefore the Feynman rules of~\cite{han1} do not apply.  The SM
fermions of the first two generation of quarks and all three
generations of leptons have their usual SM couplings.  However, in
the third generation the new Yukawa interactions required to
cancel the quadratic divergences of the top quark will shift some
SM couplings at $\mathcal{O} (v^2/f^2)$.  The $t'_{+}$ and
$t'_{-}$ couplings are also new in the T-parity littlest Higgs so
we will include all these Feynman rules in Table~\ref{newfeyntab}
and Table~\ref{newfeyngauge}.

\begin{table}[h]
\center{\begin{tabular}{|c|c||c|c|} \hline Particles & Vertices &
Particles & Vertices \\
\feyn {\bar{t} t h} {-i \frac{m_t}{v}} {\bar{t} t h h}
{i\frac{m_t(1+s_\lambda^2)}{f^2}}

\feyn {\bar{t} t\phi^0\phi^0} {i\frac{2 m_t}{3 f^2}\left(-2+3
s_\lambda^2\right)} {\bar{t}t \phi^P\phi^P} {i\frac{2 m_t}{3
f^2}\left(-2+3 s_\lambda^2\right)}

\feyn {\bar{t} t \phi^-\phi^+} {i\frac{2 m_t}{3 f^2}\left(-1+3
s_\lambda^2\right)} {\bar{t} t \phi^{--}\phi^{++}} {i \frac{2 m_t
s_\lambda^2}{f^2}}

\feyn {\bar{t'}_+ t'_+ h} {i \frac{m_t c_\lambda s_\lambda}{f}}
{\bar{t'}_+ t'_+ h h} {i \frac{m_t c_\lambda}{v f s_\lambda}}

\feyn {\bar{t'}_+ t'_+ \phi^0\phi^0} {i \frac{2 m_t c_\lambda}{v f
s_\lambda}} {\bar{t'}_+ t'_+ \phi^P\phi^P} {i \frac{2 m_t
c_\lambda}{v f s_\lambda}}

\feyn {\bar{t'}_+ t'_+ \phi^-\phi^+} {i \frac{2 m_t c_\lambda}{v f
s_\lambda}} {\bar{t'}_+ t'_+ \phi^{--}\phi^{++}} {i \frac{2 m_t
c_\lambda}{v f s_\lambda}}

\feyn {\bar{t'}_+ t h} {-i
m_t\left(\frac{s_\lambda^2}{f}P_R-\frac{c_\lambda}{s_\lambda
    v}P_L\right)}
{\bar{t'}_+ t h h} {-i \frac{m_t}{f} \left(\frac{1}{v} P_R +
\frac{(1+s_\lambda^2)
    c_\lambda}{s_\lambda f}P_L\right)}

\feyn {\bar{t'}_+ t \phi^0\phi^0} {-2 i
\frac{m_t}{f}\left(\frac{1}{v}P_R-\frac{(2-3
    s_\lambda^2)c_\lambda}{3 s_\lambda f}P_L\right)}
{\bar{t'}_+ t \phi^P\phi^P} {-2 i
\frac{m_t}{f}\left(\frac{1}{v}P_R-\frac{(2-3
    s_\lambda^2)c_\lambda}{3 s_\lambda f}P_L\right)}

\feyn {\bar{t'}_+ t \phi^-\phi^+} {-2 i
\frac{m_t}{f}\left(\frac{1}{v}P_R-\frac{(1-3
    s_\lambda^2)c_\lambda}{3 s_\lambda f}P_L\right)}
{\bar{t'}_+ t \phi^{--}\phi^{++}} {-2 i
\frac{m_t}{f}\left(\frac{1}{v}P_R+\frac{c_\lambda
    s_\lambda}{f}P_L\right)}

\feyn {\bar{t'}_- t'_+ \phi^P} {\sqrt{2} m_t \frac{c_\lambda v}{3
f^2 s_\lambda}P_R} {\bar{t'}_- t'_+ \phi^P h} {2\sqrt{2} m_t
\frac{c_\lambda}{3 f^2 s_\lambda}P_R}

\feyn {\bar{t'}_- t \phi^P} {-\sqrt{2} m_t \frac{v}{3 f^2}P_R}
{\bar{t'}_- t \phi^P h} {-\sqrt{2} m_t \frac{2}{3 f^2}P_R}

\feyn {\bar{b} t \phi^0 \phi^-} {i m_t \frac{2}{3 f^2} P_R}
{\bar{b} t \phi^P \phi^-} {- m_t \frac{2}{3 f^2} P_R}

\feyn {\bar{b} t \phi^+ \phi^{--}} {i m_t \frac{2\sqrt{2}}{3 f^2}
P_R} {\bar{b} t'_+ \phi^0 \phi^-} {-i m_t \frac{2 c_\lambda}{3
s_\lambda f^2} P_R}

\feyn {\bar{b} t'_+ \phi^P \phi^-} {m_t \frac{2 c_\lambda}{3
s_\lambda f^2} P_R} {\bar{b} t'_+ \phi^+ \phi^{--}} {- i m_t
\frac{2\sqrt{2} c_\lambda}{3 s_\lambda f^2} P_R}

\hline
\end{tabular}}
\caption{Feynman rules for the third generation quarks-scalars
which are shifted from the SM vertices, and interactions of the
$t'_{+}$ and $t'_{-}$ quarks with scalars.
$P_L=\frac{1-\gamma^5}{2}$ and $P_R=\frac{1+\gamma^5}{2}$ are the
usual LH and RH projectors.}\label{newfeyntab}
\end{table}

\begin{table}[h]
\center{\begin{tabular}{|c|c||c|c|} \hline Particles & Vertices &
Particles & Vertices \\

\feyn {\bar{t'}_- t'_- G^\mu} {i g_s\gamma_\mu} {\bar{t'}_+ t'_+
G^\mu} {i g_s\gamma_\mu}

\feyn {\ah{\mu} \tpb \tm} {\frac{2 i g'}{5} \gamma_\mu \left( P_L +
s_\lambda P_R \right) } {\ah{\mu} \tmb t} {\frac{2 i g'}{5}
c_\lambda \gamma_\mu \left( c_\lambda \EPS P_L + P_R \right) }

\feyn {\zh{\mu} \tmb t} {\frac{2}{5} i x_h g' c_\lambda \EPSs
\gamma_\mu P_R} {\zh{\mu} \tmb \tp} {\frac{2}{5} i x_h g' \EPSs
\gamma_\mu \left( P_L + s_\lambda P_R \right)}

\feyn {A^\mu \tpb \tp} {\frac{2}{3} i e \gamma_\mu} {A^\mu \tmb \tm}
{\frac{2}{3} i e \gamma_\mu}

\feyn {Z^\mu \bar{t} t} {\frac{i g}{\cw} \gamma_\mu \left(
(\frac{1}{2} - 2/3 \sw^2- \frac{1}{2} \EPSs c_\lambda^4 ) P_L -
  \frac{2}{3} \sw^2 P_R \right) }
{Z^\mu \tpb \tp} {-\frac{2}{3} \frac{i g}{\cw} \sw^2 \gamma_\mu +
  \frac{1}{2} \frac{i g}{c_w} \EPSs c_\lambda^4 P_L}

\feyn {Z^\mu \tmb \tm} {-\frac{2}{3} \frac{i g}{\cw} \sw^2
\gamma_\mu} {Z^\mu \tpb t} {-\frac{1}{2} \frac{i g}{\cw} \EPS c_\lambda^2
\gamma_\mu P_L}

\feyn {W^{+\mu} \bar{t} b} {\frac{V_{tb}}{\sqrt{2}} i g \gamma_\mu
\left( 1 - \frac{c_\lambda^4}{2} \EPSs
  \right) P_L}
{W^{+\mu} \tpb b} {-\frac{i g}{\sqrt{2}} V_{tb} \EPS c_\lambda^2
\gamma_\mu P_L}

\hline
\end{tabular}}
\caption{Feynman rules for the third generation quarks which are
shifted from the SM vertices, and interactions of the $t'_{+}$ and
$t'_{-}$ quarks.  $P_L=\frac{1-\gamma^5}{2}$ and
$P_R=\frac{1+\gamma^5}{2}$ are the usual LH and RH
projectors.}\label{newfeyngauge}
\end{table}

The phenomenology of the
littlest Higgs with T-parity could be markedly changed by the
inclusion of the additional fermions around the TeV scale.  There exist
interactions of heavy fermions with SM particles that come from
re-expressing the kinetic terms in mass eigenstates, and are
generally of the form
\begin{equation}\label{vecint}
c \bar{\psi}_{SM} \bar{\sigma}_\mu V_H^\mu \psi_H,
\end{equation}
as shown in Section~\ref{sec:model}.  These interactions
potentially can be flavor changing depending on the implementation
of the heavy mirror fermion Yukawa term~(\ref{heavyyuk}).  There
are also interactions similar to~(\ref{vecint}) which involve
heavy scalars instead of heavy vector bosons which come
from~(\ref{heavyyuk}).  Finally there are interactions of heavy
fermions with SM gauge bosons coming from (\ref{lkinsmmass}) and
(\ref{kinmirror}).  We do not include the Feynman rules for these
interactions, since they were not necessary for the sectors
studied in this paper, and they are dependent upon the
implementation (in terms of flavor) of the mirror fermion mass
terms. For the reader interested in investigating the fermion
sector of the model in more detail, one must choose an
implementation of flavor for the heavy mirror fermion
Yukawa's~(\ref{heavyyuk}) and work out the interactions.


\begin{thebibliography}{99}
{\small

\bibitem{little1}
N.~Arkani-Hamed, A.~G.~Cohen and H.~Georgi,
Phys.\ Lett.\ B {\bf 513}, 232 (2001)
{\tt [hep-ph/0105239]}.


\bibitem{HiggsPseudo}
H.~Georgi and A.~Pais,
Phys.\ Rev.\ D {\bf 10}, 539 (1974);
Phys.\ Rev.\ D {\bf 12}, 508 (1975).

\bibitem{KaplanGeorgi}
D.~B.~Kaplan and H.~Georgi,
Phys.\ Lett.\ B {\bf 136}, 183 (1984);
Phys.\ Lett.\ B {\bf 145}, 216 (1984);
D.~B.~Kaplan, H.~Georgi and S.~Dimopoulos,
Phys.\ Lett.\ B {\bf 136}, 187 (1984);
H.~Georgi, D.~B.~Kaplan and P.~Galison,
Phys.\ Lett.\ B {\bf 143}, 152 (1984);
M.~J.~Dugan, H.~Georgi and D.~B.~Kaplan,
Nucl.\ Phys.\ B {\bf 254}, 299 (1985).




\bibitem{littlest}
N.~Arkani-Hamed, A.~G.~Cohen, E.~Katz and A.~E.~Nelson,
JHEP {\bf 0207}, 034 (2002)
{\tt [hep-ph/0206021]}.


\bibitem{littlestmoose}
N.~Arkani-Hamed, A.~G.~Cohen, E.~Katz, A.~E.~Nelson, T.~Gregoire and J.~G.~Wacker,
JHEP {\bf 0208}, 021 (2002)
{\tt [hep-ph/0206020]};

\bibitem{witek}
I.~Low, W.~Skiba and D.~Smith,
Phys.\ Rev.\ D {\bf 66}, 072001 (2002) [arXiv:hep-ph/0207243].


\bibitem{martin1}
D.~E.~Kaplan and M.~Schmaltz,
JHEP {\bf 0310}, 039 (2003) [arXiv:hep-ph/0302049].

\bibitem{cust1}
S.~Chang and J.~G.~Wacker,
Phys.\ Rev.\ D {\bf 69}, 035002 (2004) [arXiv:hep-ph/0303001].

\bibitem{terning}
W.~Skiba and J.~Terning,
Phys.\ Rev.\ D {\bf 68}, 075001 (2003) [arXiv:hep-ph/0305302].

\bibitem{cust2}
S.~Chang,
JHEP {\bf 0312}, 057 (2003) [arXiv:hep-ph/0306034].

\bibitem{martin2}
M.~Schmaltz,
JHEP {\bf 0408}, 056 (2004) [arXiv:hep-ph/0407143].

\bibitem{susyproc}
P.~Meade,
arXiv:hep-ph/0402036.

\bibitem{us1}
C.~Csaki, J.~Hubisz, G.~D.~Kribs, P.~Meade and J.~Terning,
Phys.\ Rev.\ D {\bf 67}, 115002 (2003) [arXiv:hep-ph/0211124].

\bibitem{slac}
J.~L.~Hewett, F.~J.~Petriello and T.~G.~Rizzo,
JHEP {\bf 0310}, 062 (2003) [arXiv:hep-ph/0211218].

\bibitem{us2}
C.~Csaki, J.~Hubisz, G.~D.~Kribs, P.~Meade and J.~Terning,
Phys.\ Rev.\ D {\bf 68}, 035009 (2003) [arXiv:hep-ph/0303236].

\bibitem{wudka}
J.~Wudka,
arXiv:hep-ph/0307339.

\bibitem{tpar1}
H.~C.~Cheng and I.~Low,
JHEP {\bf 0309}, 051 (2003) [arXiv:hep-ph/0308199].


\bibitem{tpar2}
H.~C.~Cheng and I.~Low,
JHEP {\bf 0408}, 061 (2004) [arXiv:hep-ph/0405243].

\bibitem{tpar3}
I.~Low,
arXiv:hep-ph/0409025.

\bibitem{maxim1}
G.~Burdman, M.~Perelstein and A.~Pierce,
Phys.\ Rev.\ Lett.\  {\bf 90}, 241802 (2003) [Erratum-ibid.\  {\bf
92}, 049903 (2004)] [arXiv:hep-ph/0212228].

\bibitem{han1}
T.~Han, H.~E.~Logan, B.~McElrath and L.~T.~Wang,
Phys.\ Rev.\ D {\bf 67}, 095004 (2003) [arXiv:hep-ph/0301040].

\bibitem{maxim2}
M.~Perelstein, M.~E.~Peskin and A.~Pierce,
Phys.\ Rev.\ D {\bf 69}, 075002 (2004) [arXiv:hep-ph/0310039].

\bibitem{cwz}
S.~R.~Coleman, J.~Wess and B.~Zumino,
Phys.\ Rev.\  {\bf 177}, 2239 (1969).
C.~G.~.~Callan, S.~R.~Coleman, J.~Wess and B.~Zumino,
Phys.\ Rev.\  {\bf 177}, 2247 (1969).



\bibitem{georgi}
H.~Georgi, ``Weak Interactions And Modern Particle Theory,'' 1985.


\bibitem{andreasjay}
A.~Birkedal-Hansen and J.~G.~Wacker,
Phys.\ Rev.\ D {\bf 69}, 065022 (2004) [arXiv:hep-ph/0306161].

\bibitem{comphep}
A.~Pukhov {\it et al.},
arXiv:hep-ph/9908288.

\bibitem{maximkonst}
A.~Birkedal, K.~Matchev and M.~Perelstein,
arXiv:hep-ph/0403004.


\bibitem{wmap}
D.~N.~Spergel {\it et al.}  [WMAP Collaboration],
Astrophys.\ J.\ Suppl.\  {\bf 148}, 175 (2003)
[arXiv:astro-ph/0302209].



\bibitem{UED}
H.~C.~Cheng, K.~T.~Matchev and M.~Schmaltz,
Phys.\ Rev.\ D {\bf 66}, 056006 (2002) [arXiv:hep-ph/0205314].



\bibitem{atlas}
G.~Azuelos {\it et al.},
arXiv:hep-ph/0402037.



}

\end{thebibliography}
\end{document}